\documentclass{article}
\pdfoutput=1
\usepackage{fullpage}
\usepackage{textcomp}

\usepackage{subcaption}
\usepackage{enumerate}
\usepackage{amsmath}
\usepackage{amsfonts}
\usepackage{graphicx}
\usepackage{epstopdf}

\newtheorem{lemma}{Lemma}
\newtheorem{theorem}[lemma]{Theorem}
\newtheorem{corollary}[lemma]{Corollary}

\newtheorem{fact}{Fact}

\title{On the expected diameter, width, and complexity of a stochastic convex-hull}
\author{
  Jie Xue\footnote{Dept. of Computer Science and Engg., Univ. of Minnesota --- Twin Cities, 4-192 Keller Hall, 200 Union St. SE, Minneapolis, MN 55455, USA}\\
  \texttt{xuexx193@umn.edu}
  \and
  Yuan Li\footnotemark[1]\\
  \texttt{lixx2100@umn.edu}
  \and
  Ravi Janardan\footnotemark[1]\\
  \texttt{janardan@umn.edu}  
}
\date{}
\begin{document}

\maketitle

\begin{abstract}
We investigate several computational problems related to the stochastic convex hull (SCH).
Given a stochastic dataset consisting of $n$ points in $\mathbb{R}^d$ each of which has an existence probability, a SCH refers to the convex hull of a realization of the dataset, i.e., a random sample including each point with its existence probability.
We are interested in computing certain expected statistics of a SCH, including diameter, width, and combinatorial complexity.
For diameter, we establish the first deterministic 1.633-approximation algorithm with a time complexity polynomial in both $n$ and $d$.
For width, two approximation algorithms are provided: a deterministic $O(1)$-approximation running in $O(n^{d+1} \log n)$ time, and a fully polynomial-time randomized approximation scheme (FPRAS).
For combinatorial complexity, we propose an exact $O(n^d)$-time algorithm.
Our solutions exploit many geometric insights in Euclidean space, some of which might be of independent interest.
\end{abstract}

\section{Introduction}
As one of the most fundamental and important structures in computational geometry, the convex hull has a wide range of applications in areas as diverse as computer graphics, pattern recognition, statistics, robotics, and computer-aided design, among others.
Traditionally, the convex hull is studied on datasets whose information is known exactly.
However, in many real-world applications, due to noise and limitation of devices, the data obtained may be imprecise or not totally reliable.
In this situation, uncertain datasets (or stochastic datasets), in which the data points are allowed to have some uncertainty, can better model real data.
In recent years, there have been a considerable amount of work regarding geometric problems on stochastic datasets.
Among them, the convex hull structure under uncertainty, known as \textit{stochastic convex hull} (SCH), has received a lot of attention \cite{agarwal2014convex,li2015expected,loffler2010largest,suri2013most}.

In this paper, we revisit several problems related to SCH.
The uncertainty model to be considered is the well-known existential uncertainty model: each data point in the stochastic dataset has a certain (known) location in the space with an uncertain existence depicted by an associated existence probability (the existences of the points are assumed to be independent).
In real-world applications, the existence probability can be used to express the reliability or importance of each data point.
Given a stochastic dataset $\mathcal{S}$ in $\mathbb{R}^d$ equipped with existential uncertainty, a SCH of $\mathcal{S}$ refers to the convex hull of a realization of $\mathcal{S}$, which can be regarded as a probabilistic polytope in $\mathbb{R}^d$.
An effective way to study the behavior of a SCH is to compute the expected values of its basic statistics, which is our main focus in this paper.
Expected statistics can used to express the ``average-case'' information of a SCH, which is quite helpful for understanding a probabilistic polytope.
In this paper, we consider three basic statistics: diameter, width, and combinatorial complexity.
Informally speaking, the diameter/width of a convex hull (or convex polytope) captures how ``large'' it is, while the combinatorial complexity measures how ``complicated'' it is.
Formal definitions can be found in Sec.~\ref{sec-prelim}.
We are interested in establishing polynomial-time algorithms (both exact and approximate) for computing the expectations of these statistics for a SCH.

Note that solving such expectation computational problems is in general much more challenging than computing the statistics of a convex hull in the traditional setting where there is no uncertainty.
The main difficulty is that one has to deal with exponentially many realizations of a stochastic dataset.
For this reason, many similar problems of this type were known to be \#P-hard \cite{huang2015approximating}, while other ones usually require much higher time costs than their non-stochastic versions.
Our polynomial-time solutions exploit many geometric insights in Euclidean space, some of which might be of independent interest.

\subsection{Related work} \label{sec-rework}
Geometric computation on uncertain data has received considerable attentions in recent years.
A general introduction can be found in \cite{loffler2009data}.
Many fundamental geometric problems have been studied under uncertainty, e.g., nearest-neighbor search \cite{agarwal2013nearest,suri2014most}, minimum spanning trees \cite{kamousi2011stochastic2}, closest pair \cite{huang2015approximating,kamousi2014closest,xue2016stochastic}, range search \cite{agarwal2012range,agarwal2016range}, linear separability \cite{fink2016hyperplane,xue2016separability}, dominance relation \cite{xue2016colored}, etc.
These problems were studied either under existential uncertainty (which is used in this paper) or under locational uncertainty (where the locations of the data points are uncertain).

There have also been several papers concerning SCH \cite{agarwal2014convex,huang2014epsilon,li2015expected,loffler2010largest,suri2013most}.
We only summarize those that are strongly relevant to this paper.
Li et al. \cite{li2015expected} studied the expected computation of some basic statistics of a SCH in $\mathbb{R}^2$, e.g., area, perimeter, diameter (their results for diameter are summarized below), etc.
The results in \cite{li2015expected} are presented in a slightly different uncertainty model, but most of the algorithms also work under existential uncertainty.
Huang et al. \cite{huang2014epsilon} studied $\varepsilon$-coresets of a stochastic dataset (under both existential and locational uncertainty), which can be used to efficiently approximate the expected directional width of a SCH with respect to any given direction (see Sec.~\ref{sec-prelim} for the definition of directional width).
One should note that, although the diameter (resp., width) is defined as the largest (resp., smallest) directional width, the $\varepsilon$-coresets constructed in \cite{huang2014epsilon} cannot be used to approximate the expected diameter/width of a SCH.
The reason is simple: the direction defining the diameter/width of a SCH varies from realization to realization, and in general the largest/smallest expected directional width (over all directions) is quite different from the expected diameter/width of a SCH.

Specifically, the expected diameter of a SCH was investigated in some recent works.
Huang and Li \cite{huang2015approximating} provided an FPRAS for computing the expected farthest-pair distance of a stochastic dataset in a metric space.
This directly implies an FPRAS for computing the expected diameter of a SCH, since in Euclidean space the farthest-pair distance of a set of points is just the diameter of their convex hull.
However, an FPRAS can only obtain the desired approximation with high probability, and there seems no way to verify whether an answer obtained by the FPRAS is truly a good approximation.
Li et al. \cite{li2015expected} gave a deterministic $(2/\sqrt{3})$-approximation algorithm in $\mathbb{R}^2$, which is based on (exactly) computing the expected diameter of the stochastic smallest enclosing ball.
Although \cite{li2015expected} only considered the case in $\mathbb{R}^2$, the algorithm can be naturally extended to compute a $(\sqrt{2d}/\sqrt{d+1})$-approximation of the expected diameter of a SCH in $\mathbb{R}^d$.
Nevertheless, the runtime of this algorithm grows exponentially as $d$ increases, since computing the expected diameter of the stochastic smallest enclosing ball requires $n^{\Omega(d)}$ time \cite{jorgensen2011geometric}.
The width and combinatorial complexity of a SCH have not yet been investigated previously, to our best knowledge.

\subsection{Our results}
\textbf{Expected diameter.}
As summarized in Sec.~\ref{sec-rework}, the existing approximation algorithm for computing the expected diameter of a SCH is not polynomial-time when the dimension $d$ is not a fixed constant.
Due to this limitation, we investigate the problem without assuming $d$ is fixed.
We ask the following question: how accurately one can approximate the expected diameter in $(n,d)$-polynomial time (i.e., time polynomial in both the dataset-size $n$ and the dimension $d$)?
In this paper, we give the first algorithm which achieves a 1.633-approximation in $(n,d)$-polynomial time (Theorem~\ref{thm-diam}).
Note that computing a 2-approximation is fairly easy (see Appendix~\ref{app-2apx}).
To obtain our result, however, requires insightful new ideas and nontrivial effort.
The main ingredient of our algorithm is a notion called \textit{witness sequence}, which can well capture the diameter of a polytope using only five points, and reduces the task of handling exponentially many realizations to considering only $O(n^5)$ possible witness sequences.
\smallskip

\noindent
\textbf{Expected width.}
We study the expected-width problem in $\mathbb{R}^d$ with a fixed dimension $d$.
Two approximation algorithms are proposed for computing the expected width: a deterministic $O(1)$-approximation running in $O(n^{d+1} \log n)$ time (Theorem~\ref{thm-wid1}), and an FPRAS (Theorem~\ref{thm-wid2}).
Both the algorithms are based on a notion called \textit{witness simplex}, which is an analogue of the witness sequence in the expected-width problem.
The witness simplex captures the width of a polytope.
It allows us to ``group'' exponentially many realizations into polynomial-many groups, and thus makes polynomial-time approximations possible.
\smallskip

\noindent
\textbf{Expected combinatorial complexity.}
We study the expected-combinatorial-complexity problem in $\mathbb{R}^d$ with a fixed dimension $d$.
We provide an exact algorithm for computing the expected combinatorial complexity of a SCH in $O(n^d)$ time.
Our algorithm uses a nontrivial reduction from the problem to SCH membership probability queries, and then takes advantage of some very recent results for the latter \cite{fink2016hyperplane,xue2016separability}.
In order to complete the computation in $O(n^d)$ time, some new ideas are needed together with an observation in \cite{fink2016hyperplane}.

\subsection{Preliminaries} \label{sec-prelim}
We give the formal definitions of some basic notions used in this paper.
A \textit{stochastic dataset} in $\mathbb{R}^d$ is a pair $\mathcal{S} = (S,\pi)$ where $S$ is a set of points in $\mathbb{R}^d$ and $\pi:S \rightarrow (0,1]$ specifies the existence probability of each point in $S$.
A \textit{realization} of $\mathcal{S}$ is a random sample $R \subseteq S$ where each point $a \in S$ is sampled with probability $\pi(a)$.
A \textit{stochastic convex hull} (SCH) of $\mathcal{S}$ refers to the convex hull of a realization of $\mathcal{S}$, which can be regarded as a probabilistic polytope in $\mathbb{R}^d$.

Let $P$ be a convex polytope in $\mathbb{R}^d$. The \textit{combinatorial complexity} (or simply \textit{complexity}) of $P$, denoted by $|P|$, is defined as the total number of the faces of $P$ (the dimensions of the faces vary from 0 to $d-1$).
If $\mathbf{u}$ is a unit vector in $\mathbb{R}^d$, we define the \textit{directional width} of $P$ with respect to $\mathbf{u}$ as

\begin{equation*}
\text{wid}_\mathbf{u}(P) = \sup_{p,q \in P} \left( \langle \mathbf{u},p \rangle - \langle \mathbf{u},q \rangle \right),
\end{equation*}
\normalsize
where $\langle \cdot,\cdot \rangle$ denotes the inner product.
Let $U$ be the set of unit vectors in $\mathbb{R}^d$.
Then the \textit{diameter} of $P$ is defined as $\text{diam}(P) = \sup_{\mathbf{u} \in U} \text{wid}_\mathbf{u}(P)$, and the \textit{width} of $P$ is defined as $\text{wid}(P) = \inf_{\mathbf{u} \in U} \text{wid}_\mathbf{u}(P)$.
It is clear that the diameter of $P$ is also the distance between the farthest-pair of points in $P$.

For two points $x = (x_1,\dots,x_d)$ and $y = (y_1,\dots,y_d)$ in $\mathbb{R}^d$, we define $x \prec y$ if the $d$-tuple $(x_1,\dots,x_d)$ is smaller than the $d$-tuple $(y_1,\dots,y_d)$ in lexicographic order.
Then $\prec$ induces a (strict) total order on $\mathbb{R}^d$, called $\prec$-\textit{order}.

The approximation algorithms in this paper use \textit{relative} performance guarantees.
Formally, if $\mathit{res}$ is the exact answer of the problem, a $\delta$-approximation ($\delta \geq 1$) algorithm outputs an answer within the range $[\mathit{res}/\delta,\mathit{res}]$, where $\mathit{res}$.


\section{Approximating the expected diameter} \label{sec-diam}
Let $\mathcal{S} = (S,\pi)$ be a stochastic dataset in $\mathbb{R}^d$ ($d$ is not assumed to be fixed), and suppose $|S|=n$.
Our goal in this section is to (approximately) compute the expected diameter of a SCH of $\mathcal{S}$, defined as

\begin{equation*}
\text{diam}_\mathcal{S} = \sum_{R \subseteq S} \Pr[R] \cdot \text{diam}(\mathcal{CH}(R)),
\end{equation*}
\normalsize
where $\Pr[R]$ denotes the probability that $R$ occurs as a realization of $\mathcal{S}$.
We show in Appendix~\ref{app-hard} that computing $\text{diam}_\mathcal{S}$ exactly is \#P-hard if $d$ is not fixed.

\subsection{The witness sequence}
\begin{figure}[h]
	\centering
	\includegraphics[]{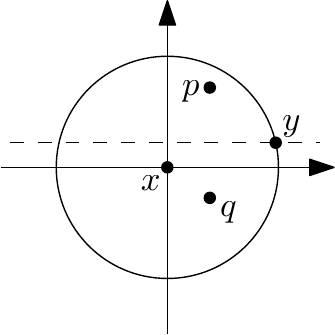}
	\caption{The locations of $x$, $p$, $q$ and $y$}
	\label{figure:cbcb1}
\end{figure}

In this section, we introduce an important notion called \textit{witness sequence} which will be used in our approximation algorithm.
Let $P$ be a convex polytope in $\mathbb{R}^d$, and $V$ be the vertex set of $P$.
For any point $x \in \mathbb{R}^d$, we define $\Phi_P(x)$ as the set of all points in $P$ farthest from $x$. Formally, $\Phi_P(x) = \{y \in P: \text{dist}(x,y) \geq \text{dist}(x,y') \text{ for any } y' \in P\}$.
Note that $\Phi_P(x) \subseteq V$, and in particular $\Phi_P(x)$ is finite.
Our first observation about $\textnormal{diam}(P)$ is the following.
\begin{lemma} \label{lem-geo1}
	Let $x \in \mathbb{R}^d$ be a point.
	If there exist $p,q \in P$ such that $\textnormal{dist}(p,q) = \textnormal{diam}(P)$ and $\angle pxq = \theta > \pi/2$, then for any $y \in \Phi_P(x)$ and $z \in \Phi_P(y)$ we have
	
	\begin{equation*}
	\textnormal{dist}(y,z) \geq \frac{\textnormal{diam}(P)}{2 \sin (\pi/2-\theta/4)}.
	\end{equation*}
	\normalsize
\end{lemma}
\textit{Proof.}
Let $x \in \mathbb{R}^d$ be a point, and suppose we have $p,q \in P$ such that $\text{dist}(p,q) = \text{diam}(P)$ and $\angle pxq > \pi/2$.
Also, let $y \in \Phi_P(x)$ be any point.
Since $\text{dist}(y,z) \geq \max\{\text{dist}(y,p), \text{dist}(y,q)\}$ for any $z \in \Phi_P(y)$, it suffices to show 

\begin{equation*}
\max\{\text{dist}(y, p), \text{dist}(y, q)\} \geq \frac{\text{diam}(P)}{2 \sin (\pi/2-\theta/4)}.
\end{equation*}
\normalsize
Without loss of generality, we may assume $x = (0,\dots,0)$, $p = (\alpha,\beta,0,\dots,0)$, $q = (\alpha,\gamma,0,\dots,0)$, where $\alpha \geq 0$ (if this is not the case, one can properly apply an isometric transformation on $\mathbb{R}^d$ to make it true).
Furthermore, we may also assume $\text{dist}(x,y)=1$, hence $\alpha^2+\beta^2 \leq 1$ and $\alpha^2+\gamma^2 \leq 1$.
Since $\angle pxq > \pi/2$, we must have $\beta \gamma < 0$ (so suppose $\beta > 0$ and $\gamma < 0$).
We first claim that $\max\{\text{dist}(y,p),\text{dist}(y,q)\}$ is minimized when

\begin{equation} \label{eq-coord}
y = \left( \sqrt{1-\frac{(\beta+\gamma)^2}{4}},\frac{\beta+\gamma}{2},0,\dots,0 \right).
\end{equation}
\normalsize
Let $y$ be the point with the above coordinates (see Figure~\ref{figure:cbcb1}), and $r = (r_1,\dots,r_d)$ be another point satisfying $\text{dist}(x,r)=1$ (i.e., $\sum_{i=1}^{d} r_i^2=1$).
First consider the case of $r_2 \leq (\beta+\gamma)/2$.
In this case, we show that $\text{dist}(r,p) \geq \max\{\text{dist}(y,p),\text{dist}(y,q)\}$.
Since $\text{dist}(y,p)=\text{dist}(y,q)$, it suffices to show $\text{dist}(r,p) \geq \text{dist}(y,p)$.
We have 

\begin{equation*}
\text{dist}^2(r,p) = 1 + \alpha^2 + \beta^2 - 2 r_1 \alpha - 2 r_2 \beta \text{ and } \text{dist}^2(y,p) = 1 + \alpha^2 + \beta^2 - 2 y_1 \alpha - 2 y_2 \beta,
\end{equation*}
\normalsize
where $y_1$ and $y_2$ are the first two coordinates of $y$ defined above.
Now we only need to show $r_1 \alpha + r_2 \beta \leq y_1 \alpha + y_2 \beta$.
Note that $r_1 \alpha + r_2 \beta \leq \alpha \sqrt{1-r_2^2} + r_2 \beta$ as $\alpha \geq 0$.
Define vectors $\mathbf{v} = (\alpha,\beta)$, $\mathbf{u} = (\sqrt{1-r_2^2},r_2)$, $\mathbf{w} = (y_1,y_2)$.
Since $\alpha \geq 0$, $y_1 > 0$, and $r_2 \leq y_2 < \beta$, the angle between $\mathbf{v}$ and $\mathbf{u}$ is greater than that between $\mathbf{v}$ and $\mathbf{w}$.
Furthermore, $\lVert \mathbf{u} \rVert_2 = \lVert \mathbf{w} \rVert_2 = 1$.
Therefore, $\alpha \sqrt{1-r_2^2} + r_2 \beta = \langle \mathbf{u},\mathbf{v} \rangle \leq \langle \mathbf{w},\mathbf{v} \rangle = y_1 \alpha + y_2 \beta$, which implies $r_1 \alpha + r_2 \beta \leq y_1 \alpha + y_2 \beta$.
In the other case $r_2 \geq (\beta+\gamma)/2$, symmetrically, we have $\text{dist}(r,q) \geq \max\{\text{dist}(y,p),\text{dist}(y,q)\}$.
Therefore, we know that $\max\{\text{dist}(y,p),\text{dist}(y,q)\}$ is minimized when $y$ has the coordinates in Equation~\ref{eq-coord}.
Note that when $y$ has these coordinates, 

\begin{equation} \label{eq-dist}
\text{dist}(y,p) = \text{dist}(y,q) = \frac{\text{dist}(p,q)}{2\sin(\angle pyq /2)} = \frac{\text{diam}(P)}{2\sin(\angle pyq /2)}.
\end{equation}
\normalsize
Next, we show that $\angle pyq \leq \pi-\theta/2$ where $\theta = \angle pxq$.
Since $\text{dist}(x,p) \leq \text{dist}(x,y)$, $\angle xyp \leq \angle xpy$.
Also, since $\text{dist}(x,q) \leq \text{dist}(x,y)$, $\angle xyq \leq \angle xqy$.
It follows that $\angle pyq = \angle xyp + \angle xyq \leq \angle xpy + \angle xqy$.
But $\angle pxq + \angle pyq + \angle xpy + \angle xqy = 2\pi$ and $\angle pxq = \theta$, which implies that $2 \angle pyq \leq 2\pi-\theta$, as desired.
Using Equation~\ref{eq-dist}, we can conclude that $\text{dist}(y,p) \geq \textnormal{diam}(P)/ (2 \sin (\pi/2-\theta/4))$, which completes the proof.
\hfill $\Box$
\medskip

\begin{figure}[h]
	\centering
	\includegraphics[]{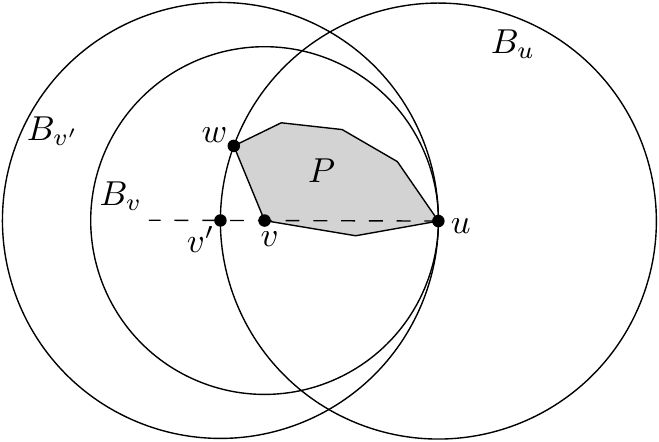}
	\caption{An illustration of $B_u,B_v,B_{v'}$}
	\label{figure:cbcb2}
\end{figure}

\noindent
Basically, Lemma~\ref{lem-geo1} states that for a point $x \in \mathbb{R}^d$, if we take $y \in P$ farthest from $x$ and $z \in P$ farthest from $y$, then the distance between $y$ and $z$ gives us a good approximation for $\textnormal{diam}(P)$ as long as there exists a pair $p,q \in P$ defining $\textnormal{diam}(P)$ with a large angle $\angle pxq$.
However, without the existence of such a pair $p,q \in P$, the approximation fails.
To handle this, we need our second observation.
\begin{lemma} \label{lem-geo2}
	Let $v \in V$ be a vertex of $P$, and $u \in \Phi_P(v), w \in \Phi_P(u)$ be two points.
	Suppose $r$ is the ray with initial point $u$ which goes through $v$, and $x$ is the point on $r$ which has distance $\textnormal{dist}(u,w)/2$ from $u$.
	Then if there exist $p,q \in P$ with $\textnormal{dist}(p,q) = \textnormal{diam}(P)$ and $\angle pxq = \theta$, we have
	
	\begin{equation*}
	\textnormal{dist}(u,w) \geq \min\left\{\textnormal{diam}(P),\frac{\textnormal{diam}(P)}{\sqrt{3}\sin(\theta/2)}\right\}.
	\end{equation*}
	\normalsize
\end{lemma}
\textit{Proof.}
Let $B_v$ be the (closed) ball centered at $u$ with radius $\text{dist}(v,u)$, and $B_u$ be the (closed) ball centered at $u$ with radius $\text{dist}(u,w)$.
Then we have $P \subseteq B_u \cap B_v$, because $u \in \Phi_P(v)$ and $w \in \Phi_P(u)$.
Now let $r$ and $x$ be the ray and the point defined in the lemma.
Define $v'$ as the point on $r$ which has distance $\text{dist}(u,w)$ from $u$, so $x$ is the midpoint of the segment connecting $v'$ and $u$.
Set $B_{v'}$ to be the (closed) ball centered at $v'$ with radius $\text{dist}(u,w)$.
See Figure~\ref{figure:cbcb2} for an illustration of the balls $B_u,B_v,B_{v'}$.
Note that $B_v \subseteq B_{v'}$, since $\text{rad}(B_{v'}) \geq \text{rad}(B_v) + \text{dist}(v,v')$ where $\text{rad}(\cdot)$ denotes the radius of a ball.
Therefore, $P \subseteq B_u \cap B_{v'}$.
Next, we claim that $B_u \cap B_{v'} \subseteq B_x$, where $B_x$ is the (closed) ball centered at $x$ with radius $\sqrt{3} \cdot \text{dist}(u,w)/2$.
Suppose $y \in B_u \cap B_{v'}$ is a point, and assume $\text{dist}(y,u) \geq \text{dist}(y,v')$ without loss of generality (so $\angle yxu \geq \pi/2$).
Define $\mu = \text{dist}(u,x)$ and $\gamma = \text{dist}(y,x)$.
Then $\gamma = \mu \cdot \sin \angle yux / \sin \angle uyx$.
Note that we have the restrictions $\angle yxu \geq \pi/2$ and $\text{dist}(u,y) \leq \text{dist}(u,v') = 2 \mu$.
Under these restrictions, it is easy to see that $\gamma$ is maximized when $\text{dist}(u,y) = 2 \mu$ and $\angle yxu = \pi/2$.
In this case, $\gamma = \sqrt{3} \mu = \text{rad}(B_x)$.
Consequently, $B_u \cap B_{v'} \subseteq B_x$, which in turn implies $P \subseteq B_x$.
With this observation, we now show the inequality in the lemma.
Let $p,q \in P \subseteq B_x$ be two points satisfying $\textnormal{dist}(p,q) = \textnormal{diam}(P)$ and $\angle pxq = \theta$.
If $\textnormal{dist}(p,q) \leq \textnormal{dist}(u,w)$, we are done, so assume $\textnormal{dist}(p,q) > \textnormal{dist}(u,w)$.
But both $\textnormal{dist}(x,p)$ and $\textnormal{dist}(x,q)$ are at most $\text{rad}(B_x) = \sqrt{3} \cdot \text{dist}(u,w)/2$.
Therefore, $\theta$ is the largest angle of the triangle $\triangle pxy$.
In this case, it is easy to see that $\textnormal{dist}(p,q)$ is maximized when $\textnormal{dist}(x,p) = \textnormal{dist}(x,q) = \text{rad}(B_x)$.
It follows that $\textnormal{dist}(p,q) \leq \sqrt{3}\sin(\theta/2) \cdot \textnormal{dist}(u,w)$, which completes the proof.
\hfill $\Box$
\medskip


\noindent
Lemma~\ref{lem-geo2} states that for a vertex $v \in V$, if we take $u \in P$ farthest from $v$ and $w \in P$ farthest from $v$, then the distance between $u$ and $w$ gives us a good approximation for $\textnormal{diam}(P)$ as long as there exists a pair $p,q \in P$ defining $\textnormal{diam}(P)$ with a small angle $\angle pxq$ (see the lemma for the definition of $x$).
The approximation is not satisfactory when $\angle pxq$ is large.
Fortunately, we already have Lemma~\ref{lem-geo1}, which is helpful for this case.
Indeed, in the case that $\angle pxq$ is large, if we further take $y \in P$ farthest from $x$ and $z \in P$ farthest from $y$, then Lemma~\ref{lem-geo1} implies that the distance between $y$ and $z$ is a good approximation for $\textnormal{diam}(P)$.
Therefore, intuitively, by taking $\max\{\text{dist}(u,w),\text{dist}(y,z)\}$, we can well-approximate $\textnormal{diam}(P)$ no matter whether $\angle pxq$ is small or large.
We formally state this as follows.
\begin{corollary} \label{cor-diam}
	Let $v,u,w,x$ be the points defined in Lemma~\ref{lem-geo2}.
	Also, let $y \in \Phi_P(x)$ and $z \in \Phi_P(y)$ be any two points.
	Then we have
	
	\begin{equation*}
	\frac{\textnormal{diam}(P)}{2\sqrt{2}/\sqrt{3}} \leq \max\{\textnormal{dist}(u,w),\textnormal{dist}(y,z)\} \leq \textnormal{diam}(P).
	\end{equation*}
	\normalsize
\end{corollary}
\textit{Proof.}
It is clear that $\max\{\text{dist}(u,w),\text{dist}(y,z)\} \leq \text{diam}(P)$, because $u,w,y,z \in P$.
Let $p,q \in P$ be two points such that $\text{dist}(p,q) = \text{diam}(P)$.
Set $\theta = \angle pxq$.
If $\theta \leq \pi/2$, then Lemma~\ref{lem-geo2} implies $\text{dist}(u,w) \geq \text{diam}(P)/(\sqrt{3}/\sqrt{2})$.
So assume $\theta > \pi/2$.
By Lemma~\ref{lem-geo1}, we have

\begin{equation*}
\text{dist}(y,z) \geq \frac{\text{diam}(P)}{2 \sin (\pi/2-\theta/4)} = \frac{\text{diam}(P)}{ 2 \cos(\theta/4)}.
\end{equation*}
\normalsize
Also, by Lemma~\ref{lem-geo2}, we have $\textnormal{dist}(u,w) \geq \text{diam}(P)/ (\sqrt{3}\sin(\theta/2))$.
Therefore,

\begin{equation*}
\max\{\text{dist}(u,w),\text{dist}(y,z)\} \geq \frac{\text{diam}(P)}{\min\{2 \cos(\theta/4), \sqrt{3}\sin(\theta/2)\}}.
\end{equation*}
\normalsize
Note that for $\theta \in (\pi/2,\pi]$, $2 \cos(\theta/4)$ is monotonically decreasing and $\sqrt{3}\sin(\theta/2)$ is monotonically increasing.
Thus, the right side of the above inequality is minimized when $2 \cos(\theta/4) = \sqrt{3}\sin(\theta/2)$.
We have this equality when $\sin(\theta/4) = 1/\sqrt{3}$, because $\sin(\theta/2) = 2 \sin(\theta/4) \cos(\theta/4)$.
By some direct calculations, we obtain the inequality in the corollary.
\hfill $\Box$
\medskip

With the five points $v,u,w,y,z$ (which are in fact the vertices of $P$) in hand, Corollary~\ref{cor-diam} allows us to approximate $\textnormal{diam}(P)$ within a factor of $2\sqrt{2}/\sqrt{3} \approx 1.633$.
In other words, the diameter information of $P$ is well ``encoded'' in those five vertices.
However, the choice of $v,u,w,y,z$ is not unique in our above construction.
For later use, we need to make it unique, which can be easily done by considering $\prec$-order (see Sec.~\ref{sec-prelim}).
We define $v \in V$ as the largest vertex of $P$ under $\prec$-order.
Also, we require $u \in \Phi_P(v)$, $w \in \Phi_P(u)$, $y \in \Phi_P(x)$, $z \in \Phi_P(y)$ to be the largest under $\prec$-order.
In this way, we obtain a uniquely defined 5-tuple $(v,u,w,y,z)$ for the polytope $P$.
We call this 5-tuple the \textit{witness sequence} of $P$, denoted by $\text{wit}(P)$.
For a 5-tuple $\psi = (x_1,\dots,x_5)$ of points in $\mathbb{R}^d$, define $\varLambda(\psi) = \max\{\text{dist}(x_2,x_3),\text{dist}(x_4,x_5)\}$.
Then Corollary~\ref{cor-diam} implies

\begin{equation} \label{ineq-diam}
\frac{\text{diam}(P)}{2\sqrt{2}/\sqrt{3}} \leq \varLambda(\text{wit}(P)) \leq \text{diam}(P)
\end{equation}
\normalsize
for any convex polytope $P$ in $\mathbb{R}^d$.

\subsection{An $(n,d)$-polynomial-time approximation algorithm} \label{sec-algdiam}
In this section, we use the notion of witness sequence defined above to establish our approximation algorithm for computing $\text{diam}_\mathcal{S}$.
Given the stochastic dataset $\mathcal{S}=(S,\pi)$, we first do a preprocessing to sort all the points in $S$ in $\prec$-order and compute the pair-wise distances of the points in $S$.
This preprocessing can be done in $O(dn^2)$ time.
Now we consider how to approximate $\text{diam}_\mathcal{S}$.
We define

\begin{equation*}
\text{diam}_\mathcal{S}^* = \sum_{R \subseteq S} \Pr[R] \cdot \varLambda(\text{wit}(\mathcal{CH}(R))).
\end{equation*}
\normalsize
Inequality~\ref{ineq-diam} implies $\text{diam}_\mathcal{S}/(2\sqrt{2}/\sqrt{3}) \leq \text{diam}_\mathcal{S}^* \leq \text{diam}_\mathcal{S}$.
Thus, in order to achieve a 1.633-approximation $\text{diam}_\mathcal{S}$, it suffices to compute $\text{diam}_\mathcal{S}^*$.
Computing $\text{diam}_\mathcal{S}^*$ by directly using the above formula takes exponential time, as $S$ has $2^n$ subsets.
However, since for any $R \subseteq S$ the witness sequence $\text{wit}(\mathcal{CH}(R))$ must be a 5-tuple of points in $S$, we can also write $\text{diam}_\mathcal{S}^*$ as

\begin{equation} \label{eq-appxdiam}
\text{diam}_\mathcal{S}^* = \sum_{\psi \in \varPsi_S} \Pr[\psi] \cdot \varLambda(\psi),
\end{equation}
\normalsize
where $\varPsi_S$ is the set of all 5-tuples of points in $S$ and $\Pr[\psi]$ is the probability that the witness sequence of a SCH of $\mathcal{S}$ is $\psi$.
Note that $|\varPsi_S| = O(n^5)$.
Thus, we can efficiently compute $\text{diam}_\mathcal{S}^*$ as long as $\Pr[\psi]$ and $\varLambda(\psi)$ can be computed efficiently for every $\psi \in \varPsi_S$.
Clearly, $\varLambda(\psi)$ can be directly computed in constant time (after our preprocessing).
To compute $\Pr[\psi]$, suppose $\psi = (p_1,\dots,p_5) \in \varPsi_S$.
It is easy to check that if $p_1 = p_2$, then either $\Pr[\psi]=0$ or $\varLambda(\psi)=0$.
So we may assume $p_1 \neq p_2$.
In this case, we give the following criterion for checking if $\psi$ is the witness sequence of a SCH of $\mathcal{S}$.
For three points $a,b,c \in \mathbb{R}^d$, we write $a \prec_b c$ if $\text{dist}(a,b) < \text{dist}(c,b)$, or $\text{dist}(a,b) = \text{dist}(c,b)$ and $a \prec c$.
\begin{lemma} \label{lem-seqcond}
	Let $\psi = (p_1,\dots,p_5) \in \varPsi_S$ with $p_1 \neq p_2$.
	Suppose $r$ is the ray with initial point $p_2$ which goes through $p_1$, and $x$ is the point on $r$ which has distance $\textnormal{dist}(p_2,p_3)/2$ from $p_2$.
	For a realization $R$ of $\mathcal{S}$, we have $\psi = \textnormal{wit}(\mathcal{CH}(R))$ iff the following two conditions hold. \\
	\textnormal{(1)} $R$ contains $p_1,\dots,p_5$. \\
	\textnormal{(2)} $R$ does not contain any point $a \in S$ satisfying $p_1 \prec a$ or $p_2 \prec_{p_1} a$ or $p_3 \prec_{p_2} a$ or $p_4 \prec_x a$ or $p_5 \prec_{p_4} a$.
\end{lemma}
\textit{Proof.}
Let $R$ be a realization of $\mathcal{S}$, and set $C = \mathcal{CH}(R)$.
The proof of the lemma is somehow straightforward by using the definition of witness sequence.
To see the ``if'' part, assume the two conditions in the lemma hold.
Then $p_1$ must be the largest point in $R$ under $\prec$-order, which must be a vertex of $C$.
Furthermore, $p_2,p_3,p_4,p_5$ must be the largest points in $\Phi_C(p_1),\Phi_C(p_2),\Phi_C(x),\Phi_C(p_4)$ under $\prec$-order, respectively.
Thus, by definition, $\psi = (p_1,\dots,p_5) = \text{wit}(C)$.
To see the ``only if'' part, assume $\text{wit}(C) = \psi$.
Then $p_1,\dots,p_5$ are vertices of $C$ and must be contained in $R$, which implies (1).
By definition, $p_1$ is the largest vertex of $C$ under $\prec$-order, and $p_2,p_3,p_4,p_5$ are the largest points in $\Phi_C(p_1),\Phi_C(p_2),\Phi_C(x),\Phi_C(p_4)$ under $\prec$-order respectively, which implies (2).
\hfill $\Box$
\medskip

\noindent
By Lemma~\ref{lem-seqcond}, it is quite easy to compute $\Pr[\psi]$ in linear time, just by multiplying the existence probabilities of the points in $\psi$ and the non-existence probabilities of all the points which should not be included in $R$ (according to the condition (2) in the lemma).
Using Equation~\ref{eq-approx}, we obtain an $(n,d)$-polynomial-time algorithm to compute $\text{diam}_\mathcal{S}^*$. 
This algorithm runs in $O(n^6 + dn^2)$ time.
But we can easily improve the runtime to $O(n^5 \log n + dn^2)$; see Appendix~\ref{app-impdiam}.
\begin{theorem} \label{thm-diam}
	One can achieve a $1.633$-approximation of $\textnormal{diam}_\mathcal{S}$ in $(n,d)$-polynomial time.
	Specifically, the approximation can be done in $O(n^5 \log n + dn^2)$ time.
\end{theorem}

Interestingly, our witness-sequence technique also gives an $O(dn)$-time 1.633-approximation algorithm for computing the diameter of the convex hull of a (non-stochastic) point-set $S$ in $\mathbb{R}^d$, because $\text{wit}(\mathcal{CH}(S))$ can be computed in $O(dn)$ time.
To our best knowledge, there has not been any linear-time algorithm which can achieve such an approximation factor when $d$ is not fixed.

\section{Approximating the expected width}
Let $\mathcal{S} = (S,\pi)$ be a stochastic dataset in $\mathbb{R}^d$ with $d$ fixed, and suppose $|S|=n$.
Our goal in this section is to (approximately) compute the expected width of a SCH of $\mathcal{S}$, defined as

\begin{equation*}
\text{wid}_\mathcal{S} = \sum_{R \subseteq S} \Pr[R] \cdot \text{wid}(\mathcal{CH}(R)),
\end{equation*}
\normalsize
where $\Pr[R]$ denotes the probability that $R$ occurs as a realization of $\mathcal{S}$.

\subsection{The witness simplex}
Recall that when solving the expected-diameter problem, we developed the notion of witness sequence, which well-captures the diameter of a polytope and satisfies (1) the total number of the possible witness sequences of a SCH is polynomial (though there are exponentially many realizations), (2) the probability of a sequence being the witness sequence of a SCH can be easily computed.
We apply this basic idea again to the expected-width problem.
To this end, we have to design some good ``witness object'' for width, which satisfies the above two conditions.
The witness object to be defined is called \textit{witness simplex}.

Let $P$ be a convex polytope in $\mathbb{R}^d$ with $\text{wid}(P)>0$, and $V$ be the vertex set of $P$.
We choose $d+1$ vertices $v_0,\dots,v_d \in V$ of $P$ inductively as follows.
Define $v_0 \in V$ as the largest vertex of $P$ under $\prec$-order.
Suppose $v_0,\dots,v_{i}$ are already defined.
Let $E_i$ be the (unique) $i$-dim hyperplane in $\mathbb{R}^d$ through $v_0,\dots,v_{i}$ (or the $i$-dim linear subspace of $\mathbb{R}^d$ spanned by $v_0,\dots,v_{i}$).
We then define $v_{i+1} \in V$ as the vertex of $P$ which has the maximum distance to $E_i$, i.e., $v_{i+1} = \arg \max_{v \in V} \text{dist}(v,E_i)$.
If there exist multiple vertices having maximum distance to $E_i$, we choose the largest one under $\prec$-order to be $v_{i+1}$.
In this way, we obtain the vertices $v_0,\dots,v_d$.
The \textit{witness simplex} $\Delta_P$ of $P$ is defined as the $d$-simplex with vertices $v_0,\dots,v_d$.
The (ordered) sequence $(v_0,\dots,v_d)$ is said to be the \textit{vertex list} of $\Delta_P$.
Note that the vertex list is determined by only $\Delta_P$ and independent of $P$.
In other words, if we only know $\Delta_P$ without knowing the original polytope $P$, we can still recover the vertex list of $\Delta_P$, just by ordering the $d+1$ vertices of $\Delta_P$ into a sequence $(v_0,\dots,v_d)$ such that $v_0$ is the largest under $\prec$-order, and each $v_{i+1}$ is the one having the maximum distance to $E_i$ (the linear subspace spanned by $v_0,\dots,v_{i}$).
A useful geometric property of the witness simplex $\Delta_P$ is that it well-captures the width of $P$.
\begin{lemma} \label{lem-witsmplx}
	Let $P$ be a convex polytope in $\mathbb{R}^d$ with $\textnormal{wid}(P)>0$, then we have $\textnormal{wid}(\Delta_P) = \Theta(\textnormal{wid}(P))$.
	The constant hidden in $\Theta(\cdot)$ could be exponential in $d$.
\end{lemma}
\textit{Proof.}
Note that $\text{wid}(\Delta_P) \leq \text{wid}(P)$ since $\Delta_P \subseteq P$.
It suffices to show that $\text{wid}(\Delta_P) = \Omega(\text{wid}(P))$.
Let $(v_0,\dots,v_d)$ be the vertex list of $\Delta_P$.
Also, let $E_i$ be the $i$-dim hyperplane in $\mathbb{R}^d$ through $v_0,\dots,v_{i}$.
Suppose each $v_i$ has the coordinates $v_i = (y_{i,1},\dots,y_{i,d})$.
Without loss of generality, we may assume that $y_{i,j}=0$ for $j>i$, that is, $v_0 = (0,\dots,0)$, $v_1 = (y_{1,1},0,\dots,0)$, $v_2 = (y_{2,1},y_{2,2},0,\dots,0)$, and so forth (if this is not the case, one can properly apply an isometric transformation on $\mathbb{R}^d$ to make it true).
With this assumption, $E_i$ is nothing but the $i$-dim linear subspace of $\mathbb{R}^d$ spanned by the axes $x_1,\dots,x_i$.
Note that $|y_{i,i}| = \text{dist}(v_i,E_{i-1}) \geq \text{dist}(v_{i+1},E_{i-1}) \geq |y_{i+1,i+1}|$.
Therefore, $|y_{1,1}| \geq \cdots \geq |y_{d,d}|$.
Furthermore, let $v \in V$ be any vertex of $P$ with coordinates $v = (z_1,\dots,z_d)$.
For every $i \in \{1,\dots,d\}$, we have that $\text{dist}(v_i,E_{i-1}) \geq \text{dist}(v,E_{i-1}) \geq |z_i|$, which implies $-|y_{i,i}| \leq z_i \leq |y_{i,i}|$.
Based on this observation, we now show that $\text{wid}(\Delta_P) \geq c \cdot \text{wid}(P)$ for some constant $c$.
It suffices to show that there exists a constant $c$ such that $\text{wid}_\mathbf{u}(\Delta_P) \geq c \cdot \text{wid}(P)$ for any unit vector $\mathbf{u} \in \mathbb{R}^d$.
We use induction to achieve this.
First, for $\mathbf{u} = (0,\dots,0,1)$, we have

\begin{equation*}
\text{wid}_\mathbf{u}(\Delta_P) = |y_{d,d}| \geq \text{wid}_\mathbf{u}(P)/2 \geq \text{wid}(P)/2,
\end{equation*}
\normalsize
because the $d$-th coordinate of any $v \in V$ has absolute value at most $|y_{d,d}|$.
It follows that $\text{wid}_\mathbf{u}(\Delta_P) \geq c_d \cdot \text{wid}(P)$ for a constant $c_d = 1/2$.
Using this as a base case, we may assume that there exists a constant $c_{i+1} \in (0,1)$ such that $\text{wid}_\mathbf{u}(\Delta_P) \geq c_{i+1} \cdot \text{wid}(P)$ for any unit vector $\mathbf{u} \in \mathbb{R}^d$ whose first $i$ coordinates are all 0.
Our goal is to find a new constant $c_i \in (0,1)$ such that $\text{wid}_\mathbf{u}(\Delta_P) \geq c_i \cdot \text{wid}(P)$ for any unit vector $\mathbf{u} \in \mathbb{R}^d$ whose first $i-1$ coordinates are all 0.
Let $\mathbf{u} = (0,\dots,0,u_i,\dots,u_d) \in \mathbb{R}^d$ be such a unit vector, and define $\mathbf{u}' = (0,\dots,0,u_{i+1}',\dots,u_d') \in \mathbb{R}^d$ as a unit vector where $u_j' = u_j/\sqrt{1-u_i^2}$ for $j \in \{i+1,\dots,d\}$.
We may assume $u_i \geq 0$ because $\text{wid}_\mathbf{u}(\Delta_P) = \text{wid}_{-\mathbf{u}}(\Delta_P)$.
Set $c_i = c_{i+1}/5$.
We verify that $\text{wid}_\mathbf{u}(\Delta_P) \geq c_i \cdot \text{wid}(P)$ by considering two cases, $u_i|y_{i,i}| \geq c_i \cdot \text{wid}(P)$ and $u_i|y_{i,i}| < c_i \cdot \text{wid}(P)$.
In the case of $u_i|y_{i,i}| \geq c_i \cdot \text{wid}(P)$, we immediately have

\begin{equation*}
\text{wid}_\mathbf{u}(\Delta_P) \geq |\langle \mathbf{u},v_i \rangle - \langle \mathbf{u},v_{i-1} \rangle| = u_i|y_{i,i}| \geq c_i \cdot \text{wid}(P).
\end{equation*}
\normalsize
In the case of $u_i|y_{i,i}| < c_i \cdot \text{wid}(P)$, we consider the unit vector $\mathbf{u}'$ defined above.
Let $\alpha,\beta \in \{0,\dots,d\}$ be indices such that $\text{wid}_{\mathbf{u}'}(\Delta_P) = \langle \mathbf{u}',v_\alpha \rangle - \langle \mathbf{u}',v_\beta \rangle$.
We claim that $\langle \mathbf{u},v_\alpha \rangle - \langle \mathbf{u},v_\beta \rangle \geq c_i \cdot \text{wid}(P)$.
First, since the $i$-th coordinates of $v_\alpha$ and $v_\beta$ have absolute values at most $|y_{i,i}|$ (as observed before), we have

\begin{equation*}
\langle \mathbf{u},v_\alpha \rangle - \langle \mathbf{u},v_\beta \rangle \geq \sqrt{1-u_i^2} \cdot \text{wid}_{\mathbf{u}'}(\Delta_P) - 2 u_i |y_{i,i}|.
\end{equation*}
\normalsize
We have $u_i|y_{i,i}| < c_i \cdot \text{wid}(P) = (c_{i+1}/5) \cdot \text{wid}(P)$ by assumption and $\text{wid}_{\mathbf{u}'}(\Delta_P) \geq c_{i+1} \cdot \text{wid}(P)$ by the induction hypothesis, hence $\langle \mathbf{u},v_\alpha \rangle - \langle \mathbf{u},v_\beta \rangle \geq (\sqrt{1-u_i^2} - 2/5) \cdot c_{i+1} \cdot \text{wid}(P)$.
It is sufficient to show that $\sqrt{1-u_i^2} \geq 3/5$.
Note that $|y_{i,i}| \geq \text{wid}_\mathbf{w}(P)/2 \geq \text{wid}(P)/2$, where $\mathbf{w}$ is the unit vector whose $i$-th coordinate is 1 and other coordinates are 0.
By applying this to the inequality $u_i|y_{i,i}| < c_i \cdot \text{wid}(P)$, we have $u_i < 2 c_i = (2/5) \cdot c_{i+1} \leq 2/5$.
Therefore, $\sqrt{1-u_i^2} \geq 1-u_i \geq 3/5$, as we desire.
In both of the cases, we have $\langle \mathbf{u},v_\alpha \rangle - \langle \mathbf{u},v_\beta \rangle \geq c_i \cdot \text{wid}(P)$.
Since $\text{wid}_\mathbf{u}(\Delta_P) \geq \langle \mathbf{u},v_\alpha \rangle - \langle \mathbf{u},v_\beta \rangle$, it holds that $\text{wid}_\mathbf{u}(\Delta_P) \geq c_i \cdot \text{wid}(P)$.
We can use this induction argument to finally obtain the constant $c_1$ (note that $c_1$ is truly a constant as $d$ is fixed), which satisfies $\text{wid}_\mathbf{u}(\Delta_P) \geq c_1 \cdot \text{wid}(P)$ for any unit vector $\mathbf{u} \in \mathbb{R}^d$.
As a result, $\text{wid}(\Delta_P) \geq c_1 \cdot \text{wid}(P)$, completing the proof.
\hfill $\Box$

\subsection{An $O(1)$-approximation algorithm} \label{sec-algwid}
With the notion of witness simplex in hand, we propose a $O(1)$-approximation algorithm for computing $\text{wid}_\mathcal{S}$. 
The basic idea is similar to what we use for approximating $\text{diam}_\mathcal{S}$.
We define

\begin{equation*}
\text{wid}_\mathcal{S}^* = \sum_{R \subseteq S} \Pr[R] \cdot \text{wid}(\Delta_{\mathcal{CH}(R)}),
\end{equation*}
\normalsize
Lemma~\ref{lem-witsmplx} implies $\text{wid}_\mathcal{S}^* = \Theta(\text{wid}_\mathcal{S})$.
Thus, in order to approximate $\text{wid}_\mathcal{S}$ within a constant factor, it suffices to compute $\text{wid}_\mathcal{S}^*$.
To compute $\text{wid}_\mathcal{S}^*$ by directly using the above formula takes exponential time, as $S$ has $2^n$ subsets.
However, since $\Delta_{\mathcal{CH}(R)}$ must be a $d$-simplex with vertices in $S$, $\text{wid}_\mathcal{S}^*$ can also be written as

\begin{equation}
\text{wid}_\mathcal{S}^* = \sum_{\Delta \in \varGamma_S^d} \Pr[\Delta] \cdot \text{wid}(\Delta),
\end{equation}
\normalsize
where $\varGamma_S^d$ is the set of all $d$-simplices in $\mathbb{R}^d$ whose vertices are (distinct) points in $S$ and $\Pr[\Delta]$ is the probability that the witness simplex of a SCH of $\mathcal{S}$ is $\Delta$.
Note that $|\varGamma_S^d| = O(n^{d+1})$, which is polynomial.
So the above formula allows us to compute $\text{wid}_\mathcal{S}^*$ in polynomial time, as long as we are able to compute $\Pr[\Delta]$ efficiently for each $\Delta \in \varGamma_S^d$.
Fixing $\Delta \in \varGamma_S^d$, we now investigate how to compute $\Pr[\Delta]$.
As argued before, we can recover the vertex list $(v_0,\dots,v_d)$ of $\Delta$.
By the construction of $\Delta$, $v_0,\dots,v_d$ are points in $S$.
For $i \in \{0,\dots,d-1\}$, we denote by $E_i$ the $i$-dim hyperplane in $\mathbb{R}^d$ through $v_0,\dots,v_{i}$.
We give the following criterion for checking if $\Delta$ is the witness simplex of a SCH of $\mathcal{S}$.
For a hyperplane $H$ (of any dimension) in $\mathbb{R}^d$ and two points $a,b \in \mathbb{R}^d$, we write $a \prec_H b$ if $\text{dist}(a,H)<\text{dist}(b,H)$, or $\text{dist}(a,H)=\text{dist}(b,H)$ and $a \prec b$.
\begin{lemma} \label{lem-conditions}
	For a realization $R$ of $\mathcal{S}$, $\Delta$ is the witness simplex of $\mathcal{CH}(R)$ \textnormal{(}i.e., $\Delta = \Delta_{\mathcal{CH}(R)}$\textnormal{)} iff the following two conditions hold. \\
	\textnormal{(1)} $R$ contains $v_0,\dots,v_d$. \\
	\textnormal{(2)} $R$ does not contain any point $a \in S$ satisfying $v_0 \prec a$ or $v_{i+1} \prec_{E_i} a$ for some $i \in \{0,\dots,d-1\}$.
\end{lemma}
\textit{Proof.}
Let $R$ be a realization of $\mathcal{S}$, and set $C = \mathcal{CH}(R)$.
The proof of the lemma is somehow straightforward by using the definition of witness simplex.
To see the ``if'' part, assume the two conditions in the lemma hold.
Then $v_0$ must be the largest point in $R$ under $\prec$-order, which must be a vertex of $C$.
Furthermore, $v_{i+1}$ must be a vertex of $C$ (for it is the farthest from $E_i$ and the points in $S$ are in general position) which has the maximum distance to $E_i$ (in addition, if there exists another vertex $v$ of $C$ having the same distance to $E_i$ as $v_{i+1}$, then $v \prec v_{i+1}$).
Thus, by definition, $\Delta = \Delta_C$.
To see the ``only if'' part, assume $\Delta = \Delta_C$.
Then $v_0,\dots,v_d$ are vertices of $C$ and must be contained in $R$, which implies (1).
Since $(v_0,\dots,v_d)$ is the vertex list of $\Delta$, $v_0$ is the largest vertex of $C$ under $\prec$-order.
Also, for any $i \in \{0,\dots,d-1\}$, $v_{i+1}$ is a vertex of $C$ which has the maximum distance to $E_i$ (in addition, if there exists another vertex $v$ of $C$ having the same distance to $E_i$ as $v_{i+1}$, then $v \prec v_{i+1}$), so $R$ cannot contain any point $a$ with $v_{i+1} \prec_{E_i} a$.
So we have the condition (2).
\hfill $\Box$
\medskip

\noindent
Using the above lemma, we can straightforwardly compute $\Pr[\Delta]$ in linear time, just by multiplying the existence probabilities of $v_0,\dots,v_d$ and the non-existence probabilities of all $a \in S$ which should not be included in $R$ (according to the condition (2) in the lemma).
Therefore, we obtain an $O(n^{d+2})$-time algorithm for computing $\text{wid}_\mathcal{S}^*$.
It is easy to improve the runtime to $O(n^{d+1} \log n)$; see Appendix~\ref{app-impwid}.
\begin{theorem} \label{thm-wid1}
	One can $O(1)$-approximate $\textnormal{wid}_\mathcal{S}$ in $O(n^{d+1} \log n)$ time.
	The constant approximation factor could be exponential in $d$.
\end{theorem}

\subsection{A fully polynomial-time randomized approximation scheme}
In this section, we develop a fully polynomial-time randomized approximation scheme (FPRAS) for computing $\text{wid}_\mathcal{S}$.
An FPRAS should take $\mathcal{S}$ and a real number $\varepsilon>0$ as input, and output a $(1+\varepsilon)$ approximation of $\text{wid}_\mathcal{S}$ in time polynomial in the size of $\mathcal{S}$ and $1/\varepsilon$ with probability at least 2/3.

We first introduce some notations.
As defined in the preceding section, $\varGamma_S^d$ is the set of all $d$-simplices in $\mathbb{R}^d$ whose vertices are (distinct) points in $S$, and for each $\Delta \in \varGamma_S^d$ the notation $\Pr[\Delta]$ denotes the probability that the witness simplex of a SCH of $\mathcal{S}$ is $\Delta$.
Let $R$ be a realization of $\mathcal{S}$ and $\Delta \in \varGamma_S^d$ be a simplex.
From Lemma~\ref{lem-conditions}, we know that $\Delta = \Delta_{\mathcal{CH}(R)}$ iff $R$ contains the vertices of $\Delta$ but does not contain some other points in $S$ according to (2) in the lemma.
We now use $V_\Delta$ to denote the set of the vertices of $\Delta$, $X_\Delta$ to denote the set of the points in $S$ that $R$ must not contain if $\Delta = \Delta_{\mathcal{CH}(R)}$.
Let $F_\Delta = S \backslash (V_\Delta \cup X_\Delta)$, which is the set of the points in $S$ whose presence/absence in $R$ does not influence whether $\Delta = \Delta_{\mathcal{CH}(R)}$.
Define $\mathcal{F}_\Delta$ as the sub-dataset of $\mathcal{S}$ with the point-set $F_\Delta$.
Our FPRAS works as follows.
First, for each $\Delta \in \varGamma_S^d$, we randomly generate $m = \gamma \log n / \varepsilon^2$ realizations of $\mathcal{F}_\Delta$, where $\gamma$ is a large enough constant to be determined.
Let $R_1^{\Delta},\dots,R_m^{\Delta}$ be the generated realizations of $\mathcal{F}_\Delta$, and set $T_i^{\Delta} = R_i^{\Delta} \cup V_{\Delta}$.
Note that the witness simplex of $\mathcal{CH}(T_i^{\Delta})$ is $\Delta$ by Lemma~\ref{lem-conditions}.
We then compute

\begin{equation} \label{eq-approx}
\text{wid}_\mathcal{S}' = \sum_{\Delta \in \varGamma_S^d} \Pr[\Delta] \cdot \left( \sum_{i=1}^{m} \frac{\text{wid}(\mathcal{CH}(T_i^{\Delta}))}{m} \right),
\end{equation}
\normalsize
and output $\text{wid}_\mathcal{S}'$ as the approximation of $\text{wid}_\mathcal{S}$.

Next, we discuss the choice of the constant $\gamma$ and verify the correctness of our FPRAS.
By Lemma~\ref{lem-witsmplx}, we can find positive constants $k_1,k_2$ such that $k_1 \cdot \text{wid}(\Delta_P) \leq \text{wid}(P) \leq k_2 \cdot \text{wid}(\Delta_P)$ for any convex polytope $P$ in $\mathbb{R}^d$ with $\text{wid}(P)>0$.
We set $\gamma = d (k_2/k_1)^2$.
With this choice of $\gamma$, we claim the following, which shows the correctness of our FPRAS.
\begin{lemma} \label{lem-fpras}
	$(1-\varepsilon) \textnormal{wid}_\mathcal{S} \leq \textnormal{wid}_\mathcal{S}' \leq (1+\varepsilon) \textnormal{wid}_\mathcal{S}$ with probability at least $2/3$.
\end{lemma}
\textit{Proof.}
Indeed, we can write

\begin{equation*}
\text{wid}_\mathcal{S} = \sum_{\Delta \in \varGamma_S^d} \Pr[\Delta] \cdot \mathbf{E}_\Delta,
\end{equation*}
\normalsize
where $\mathbf{E}_\Delta$ is the conditional expected width of a SCH of $\mathcal{S}$ under the condition that the witness simplex of the SCH is $\Delta$.
Since $\text{wid}_\mathcal{S}'$ is computed using Equation~\ref{eq-approx}, it suffices to show that

\begin{equation} \label{eq-bound}
(1-\varepsilon) \mathbf{E}_\Delta \leq \sum_{i=1}^{m} \frac{\text{wid}(\mathcal{CH}(T_i^{\Delta}))}{m} \leq (1+\varepsilon) \mathbf{E}_\Delta
\end{equation}
\normalsize
for all $\Delta \in \varGamma_S^d$ with probability at least 2/3.
Fixing $\Delta \in \varGamma_S^d$, we can regard $\text{wid}(\mathcal{CH}(T_1^{\Delta})),\dots,\text{wid}(\mathcal{CH}(T_m^{\Delta}))$ as i.i.d. random variables.
By Lemma~\ref{lem-conditions} and the construction of each $T_i^{\Delta}$, we know that the expectation of $\text{wid}(\mathcal{CH}(T_i^{\Delta}))$ is $\mathbf{E}_\Delta$.
Furthermore, we have $k_1 \cdot \text{wid}(\Delta) \leq \text{wid}(\mathcal{CH}(T_i^{\Delta})) \leq k_2 \cdot \text{wid}(\Delta)$, since the witness simplex of $\mathcal{CH}(T_i^{\Delta})$ is $\Delta$ as argued before.
Based on these observations, we can apply Hoeffding's inequality to obtain

\begin{equation*}
\Pr\left[ \ \left| \sum_{i=1}^{m} \frac{\text{wid}(\mathcal{CH}(T_i^{\Delta}))}{m} - \mathbf{E}_\Delta \right| \geq \varepsilon \mathbf{E}_\Delta \right] \leq 2 \exp \left( -\frac{2m \cdot (\varepsilon \mathbf{E}_\Delta)^2}{(k_2-k_1)^2 \cdot \text{wid}(\Delta)^2} \right).
\end{equation*}
\normalsize
Note that $m = \gamma \log n / \varepsilon^2 = d (k_2/k_1)^2 \log n / \varepsilon^2$.
Therefore,

\begin{equation*}
-\frac{2m \cdot (\varepsilon \mathbf{E}_\Delta)^2}{(k_2-k_1)^2 \cdot \text{wid}(\Delta)^2} \leq -2d \log n,
\end{equation*}
\normalsize
since $\mathbf{E}_\Delta \geq k_1 \cdot \text{wid}(\Delta)$.
It follows that Equation~\ref{eq-bound} fails with probability $O(n^{-2d})$ for a specific $\Delta$.
Therefore, by union bound, Equation~\ref{eq-bound} holds for all $\Delta \in \varGamma_S^d$ with probability $1-O(n^{-d+1})$, which is greater than 2/3 for large $n$ (assume $d \geq 2$).
As a result, the inequality in the theorem is proved.
\hfill $\Box$
\begin{theorem} \label{thm-wid2}
	There exists an FPRAS for computing $\textnormal{wid}_\mathcal{S}$.
\end{theorem}


\section{Computing the expected combinatorial complexity}
Let $\mathcal{S} = (S,\pi)$ be a stochastic dataset in $\mathbb{R}^d$ with $d$ fixed, and suppose $|S|=n$.
Our goal in this section is to compute the expected complexity of a SCH of $\mathcal{S}$, defined as

\begin{equation*}
\text{comp}_\mathcal{S} = \sum_{R \subseteq S} \Pr[R] \cdot |\mathcal{CH}(R)|,
\end{equation*}
\normalsize
where $\Pr[R]$ denotes the probability that $R$ occurs as a realization of $\mathcal{S}$.

\subsection{Reduction to SCH membership probability queries}
Given a stochastic dataset $\mathcal{T}$ in $\mathbb{R}^d$ and a query point $q \in \mathbb{R}^d$, the SCH membership probability (of $q$ with respect to $\mathcal{T}$) refers to the probability that $q$ lies in a SCH of $\mathcal{T}$, which we denote by $\text{mem}_\mathcal{T}(q)$.
It is known that $\text{mem}_\mathcal{T}(q)$ can be computed in $O(m^{d-1})$ time for $d \geq 3$ \cite{fink2016hyperplane,xue2016separability} and $O(m \log m)$ time for $d \in \{1,2\}$ \cite{agarwal2014convex}, where $m$ is the number of the stochastic points in $\mathcal{T}$.

In this section, we reduce our problem of computing $\text{comp}_\mathcal{S}$ to SCH membership probability queries.
Let $R$ be a realization of $\mathcal{S}$.
It is clear that the faces of $\mathcal{CH}(R)$ must be simplices with vertices in $S$.
Therefore, we can rewrite the formula for $\text{comp}_\mathcal{S}$ as

\begin{equation} \label{eq-comp}
\text{comp}_\mathcal{S} = \sum_{R \subseteq S} \Pr[R] \cdot \left( \sum_{\Delta \in \varGamma_S} \sigma(R,\Delta) \right) = \sum_{\Delta \in \varGamma_S} F_\Delta,
\end{equation}
\normalsize
where $\varGamma_S$ is the set of all simplices (of dimension less than $d$) with vertices in $S$, $\sigma$ is a indicating function such that $\sigma(R,\Delta) = 1$ if $\Delta$ is a face of $\mathcal{CH}(R)$ and $\sigma(R,\Delta) = 0$ otherwise, $F_\Delta$ is the probability that $\Delta$ is a face of a SCH of $\mathcal{S}$.
We now show that for each $\Delta \in \varGamma_S$, the computation of $F_\Delta$ can be reduced to a SCH membership probability query.
Suppose $Y$ is a set of $m$ ($m \geq d+1$) points in $\mathbb{R}^d$ in general position.
Let $y_0,\dots,y_k \in Y$ be $k+1$ points where $0 \leq k \leq d-1$, and $\Delta$ be the $k$-simplex with vertices $y_0,\dots,y_k$.
Define vectors $\mathbf{u}_i = y_i - y_0$ for $i \in \{1,\dots,k\}$.
By the general position assumption, $\mathbf{u}_1,\dots,\mathbf{u}_k$ generate a $k$-dim linear subspace $H$ of $\mathbb{R}^d$.
Set $H^*$ to be the orthogonal complement of $H$ in $\mathbb{R}^d$, which is by definition the $(d-k)$-dim linear subspace of $\mathbb{R}^d$ orthogonal to $H$.
We then orthogonally project the points in $Y$ to $H^*$, and denote the set of the projection images by $Y^*$.
Note that $y_0,\dots,y_k$ are clearly projected to the same point in $H^*$, which we denote by $\hat{y}$.
We have the following observation.
\begin{lemma} \label{lem-face}
	$\Delta$ is a face of $\mathcal{CH}(Y)$ iff $\hat{y}$ is a vertex of $\mathcal{CH}(Y^*)$ in $H^*$.
\end{lemma}
\textit{Proof.}
Suppose $Y = \{y_0,y_1,\dots,y_m\}$, and let $P = \mathcal{CH}(Y)$, $P^* = \mathcal{CH}(Y^*)$.
Then any point $x \in P$ can be represented as a linear combination $x = \sum_{i=0}^{m} w_i \cdot y_i$ where $w_i \geq 0$ and $\sum_{i=0}^{m}w_i=1$, which we call \textit{convex representation}.
It is easy to check that $x$ is on the boundary of $P$ iff $x$ has a unique convex representation and in which there are at most $d$ nonzero $w_i$'s.
We first show the ``if'' part.
Assume $\Delta$ is not a face of $\mathcal{CH}(Y)$.
Then there must exist $x \in \Delta$ which is not on the boundary of $P$.
Since $\Delta$ is a simplex, there is a unique convex representation of $x$ satisfying $w_i = 0$ for all $i>k$.
But this should not be the only convex representation of $x$, because $x$ is not on the boundary of $P$.
Therefore, $x$ has another convex representation with $w_i>0$ for some $i>k$ (without loss of generality, assume $w_m>0$).
Let $\rho:\mathbb{R}^d \rightarrow H^*$ be the orthogonal projection map.
We have

\begin{equation*}
\hat{y} = \rho(x) = \rho\left( \sum_{i=0}^{m} w_i \cdot y_i \right) = \sum_{i=0}^{m} w_i \cdot \rho(y_i).
\end{equation*}
\normalsize
Note that all $\rho(y_i)$ are points in $P^*$.
Furthermore, by general position assumption, $\rho(y_m) \neq \hat{y}$.
Therefore, $\hat{y}$ is not a vertex of $P^*$.
Next, we consider the ``only if'' part.
Assume $\hat{y}$ is not a vertex of $P^*$.
Then we have $P^* = \mathcal{CH}(Y^* \backslash \{\hat{y}\})$.
It follows that $\hat{y}$ has an convex representation $\hat{y} = \sum_{i=0}^{m} w_i \cdot \rho(y_i)$ with $w_0 = \cdots = w_k = 0$.
Lifting this representation, we obtain a point $x = \sum_{i=0}^{m} w_i \cdot y_i \in P$.
Since $\rho(x) = \hat{y}$, $x$ is in the $k$-dim hyperplane $L$ spanned by $y_0,\dots,y_k$.
Now assume $\Delta$ is a face of $P$, so we must have $L \cap P = \Delta$, which implies $x \in \Delta$.
This means that $x$ has an convex representation with $w_{k+1} = \cdots = w_m = 0$.
Since $x$ has two different convex representations, it is not on the boundary of $P$, contradicting that $x \in \Delta$.
As a result, $\Delta$ is not a face of $P$.
\hfill $\Box$
\medskip

By the above lemma, we can reduce the computation of $F_\Delta$ for any $\Delta \in \varGamma_S$ to a SCH membership query as follows.
For each $i \in \{0,\dots,d-1\}$, let $\varGamma_S^i \subseteq \varGamma_S$ be the subset consisting of all $i$-simplices in $\varGamma_S$ (then $\varGamma_S = \bigcup_{i=0}^{d-1} \varGamma_S^i$).
Suppose $\Delta \in \varGamma_S^k$ is a $k$-simplex with vertices $v_0,\dots,v_k \in S$.
As before, we define vectors $\mathbf{u}_i = v_i - v_0$ for $i \in \{1,\dots,k\}$.
Then $\mathbf{u}_1,\dots,\mathbf{u}_k$ generate a $k$-dim linear subspace $H$ of $\mathbb{R}^d$, and set $H^*$ to be the orthogonal complement of $H$ in $\mathbb{R}^d$.
Let $\rho:\mathbb{R}^d \rightarrow H^*$ be the orthogonal projection map.
We define a multi-set $S' = \{\rho(a): a \in S \backslash \{v_0,\dots,v_k\}\}$ of points in $H^*$, which in turn gives us a stochastic dataset $\mathcal{S}' = (S',\pi')$ in $H^*$ where $\pi'(\rho(a)) = \pi(a)$.
Set $q = \rho(v_0) = \cdots = \rho(v_k)$.
\begin{corollary} \label{cor-fdelta}
	$F_\Delta = \prod_{i=0}^{k} \pi(v_i) \cdot (1-\textnormal{mem}_{\mathcal{S}'}(q))$.
\end{corollary}
\textit{Proof.}
Let $R$ be a realization of $\mathcal{S}$.
If $\Delta$ is a face of $\mathcal{CH}(R)$, then $v_0,\dots,v_k$ must be contained in $R$.
Furthermore, by Lemma~\ref{lem-face}, $q$ must be a vertex of the projection image of $\mathcal{CH}(R)$ in $H^*$.
By the general position assumption, this is equivalent to saying that $q$ is outside the projection image of $\mathcal{CH}(R \backslash \{v_0,\dots,v_k\})$.
Conversely, if $v_0,\dots,v_k$ are contained in $R$ and $q$ is outside the projection image of $\mathcal{CH}(R \backslash \{v_0,\dots,v_k\})$, then $\Delta$ is a face of $\mathcal{CH}(R)$ by Lemma~\ref{lem-face}.
The probability that $R$ contains $v_0,\dots,v_k$ is $\prod_{i=0}^{k} \pi(v_i)$, and the probability that $q$ is outside the projection image of $\mathcal{CH}(R \backslash \{v_0,\dots,v_k\})$ is $1-\textnormal{mem}_{\mathcal{S}'}(q)$.
These two events are clearly independent.
Therefore, we have the formula in the corollary.
\hfill $\Box$ 
\medskip

Since $H^*$ is linearly homeomorphic to $\mathbb{R}^{d-k}$, computing $\textnormal{mem}_{\mathcal{S}'}(q)$ is nothing but answering a SCH membership probability query in $\mathbb{R}^{d-k}$.
Therefore, using the algorithms for answering SCH membership probability queries \cite{fink2016hyperplane,xue2016separability}, $F_\Delta$ can be computed in $O(n^{d-k-1})$ time if $k \in \{0,\dots,d-3\}$.
Note that $|\varGamma_S^k| = O(n^{k+1})$, so we can compute the sum $\sum_{i=0}^{d-3} \sum_{\Delta \in \varGamma_S^i} F_\Delta$ in $O(n^d)$ time.
In order to further compute $\text{comp}_\mathcal{S}$ by Equation~\ref{eq-comp}, we now only need to compute $\sum_{\Delta \in \varGamma_S^{d-2}} F_\Delta$ and $\sum_{\Delta \in \varGamma_S^{d-1}} F_\Delta$.
Answering SCH membership probability queries in $\mathbb{R}^1$ and $\mathbb{R}^2$ requires $O(m \log m)$ time \cite{agarwal2014convex} (where $m$ is the size of the given stochastic dataset).
Thus, if we use the algorithm in \cite{agarwal2014convex} to calculate SCH membership probabilities, our computation task cannot be done in $O(n^d)$ time.
The next section discusses how to handle this issue.

\subsection{Handling $k=d-2$ and $k=d-1$} \label{sec-special}
Set $\lambda_1 = \sum_{\Delta \in \varGamma_S^{d-1}} F_\Delta$ and $\lambda_2 = \sum_{\Delta \in \varGamma_S^{d-2}} F_\Delta$.
For simplicity of exposition, we first fix a point $o \in \mathbb{R}^d$ such that $S \cup \{o\}$ is in general position.
For every hyperplane $E$ with $o \notin E$, we denote by $E^+$ the connected component of $\mathbb{R}^d \backslash E$ containing $o$, and by $E^-$
the other one.
Define the $\mathcal{S}$-\textit{statistic} of $E$ as a 3-tuple $\text{stat}_\mathcal{S}(E) = (p^+,p^-,A)$ where $p^+ = \prod_{a \in S \cap E^+} (1-\pi(a))$, $p^- = \prod_{a \in S \cap E^-} (1-\pi(a))$, $A = S \cap E$.
Let $\mathcal{E}$ be the collection of the hyperplanes in $\mathbb{R}^d$ which go through exactly $d$ points in $S$.
Since $S \cup \{o\}$ is in general position, $\text{stat}(E)$ is defined for every $E \in \mathcal{E}$.
We say an algorithm computes the $\mathcal{S}$-statistics for $\mathcal{E}$ if it reports $\text{stat}_\mathcal{S}(E)$ for all $E \in \mathcal{E}$ in an arbitrary order (without repetition).
\begin{lemma} \label{lem-stat1}
	If there exists an algorithm computing the $\mathcal{S}$-statistics for $\mathcal{E}$ in $O(t(n))$ time and $O(s(n))$ space, then one can compute $\lambda_1$ and $\lambda_2$ in $O(t(n))$ time and $O(s(n))$ space.
\end{lemma}
\textit{Proof.}
We first consider the computation of $\lambda_1$.
Let $\Delta \in \varGamma_S^{d-1}$ and $E \in \mathcal{E}$ be the hyperplane through the $d$ vertices of $\Delta$.
Suppose $q$ and $\mathcal{S}'$ are the point and the stochastic dataset defined in Corollary~\ref{cor-fdelta} for computing $F_\Delta$.
Since $\text{mem}_{\mathcal{S}'}(q)$ is a SCH membership query in $\mathbb{R}^1$, it is clear that $1-\text{mem}_{\mathcal{S}'}(q) = p^+ + p^- - p^+ p^-$ if $\text{stat}(E) = (p^+,p^-,A)$.
Hence $F_\Delta$ can be computed from $\text{stat}_\mathcal{S}(E)$ in constant time.
Consider the algorithm provided for computing the $\mathcal{S}$-statistics for $\mathcal{E}$.
At every time it reports some $\text{stat}_\mathcal{S}(E) = (p^+,p^-,A)$, we use it to compute the corresponding $F_\Delta$ (note that $\Delta$ can be recovered from $A$) in constant time.
By summing up all $F_\Delta$, we obtain $\lambda_1$, which is done in $O(t(n))$ time and $O(s(n))$ space.
To consider $\lambda_2$, we need a careful analysis of the witness-edge method in \cite{agarwal2014convex} for computing SCH membership probability in $\mathbb{R}^2$.
Let $\mathcal{T} = (T,\tau)$ be a stochastic dataset in $\mathbb{R}^2$, and $q \in \mathbb{R}^2$ be a query point.
The witness-edge method computes $1-\text{mem}_\mathcal{T}(q)$ as a summation of which the summands one-to-one correspond to the hyperplanes (i.e., lines) which go through $q$ and one point in $T$.
Furthermore, the summand corresponding to a hyperplane $E$ can be computed from $\text{stat}_\mathcal{T}(E)$ in constant time.
See \cite{agarwal2014convex} for the details.
Now we consider the computation of $\lambda_2$.
Let $\Delta \in \varGamma_S^{d-2}$.
Suppose $q$ and $\mathcal{S}'$ are the point and the stochastic dataset defined in Corollary~\ref{cor-fdelta} for computing $F_\Delta$.
We can regard $(\mathcal{S}',q)$ as a SCH membership probability query in $\mathbb{R}^2$.
Thus, by our observation about the witness-edge method and Corollary~\ref{cor-fdelta}, $F_\Delta$ can be expressed as a summation with summands one-to-one corresponding to the lines through $q$ and one point in the point-set of $\mathcal{S}'$ (we denote by $\mathcal{L}$ the collection of these lines).
Note that there is also an one-to-one correspondence between $\mathcal{L}$ and a sub-collection $\mathcal{E}_\Delta \subset \mathcal{E}$ containing the hyperplanes through all the $d-1$ vertices of $\Delta$.
Moreover, $\text{stat}_{\mathcal{S}'}(L)$ for $L \in \mathcal{L}$ can be recovered from $\text{stat}_\mathcal{S}(E)$ for $E \in \mathcal{E}_\Delta$ corresponding to $L$ in constant time.
Therefore, we may charge each summand of $F_\Delta$ to the corresponding hyperplane $E \in \mathcal{E}_\Delta$.
Now consider the algorithm provided for computing the $\mathcal{S}$-statistics for $\mathcal{E}$.
At every time it reports $\text{stat}_\mathcal{S}(E)$ for some $E \in \mathcal{E}$, we use it to compute all summands charged to $E$.
Note that each $E \in \mathcal{E}$ belongs to exactly $d-1$ $\mathcal{E}_\Delta$'s, and hence is charged with exactly $d-1$ summands.
Therefore, this computation can be done in constant time.
By summing up all summands charged to all $E \in \mathcal{E}$, we finally obtain $\lambda_2$, which is done in $O(t(n))$ time and $O(s(n))$ space.
\hfill $\Box$
\medskip

By the above lemma, it is now sufficient to establish a good algorithm computing the $\mathcal{S}$-statistics for $\mathcal{E}$.
We give in Appendix~\ref{app-reptstat} an algorithm computing the $\mathcal{S}$-statistics for $\mathcal{E}$ in $O(n^d)$ time and $O(n)$ space (the basic idea is implicitly known in \cite{fink2016hyperplane,xue2016separability}).
With this algorithm in hand, Lemma~\ref{lem-stat1} implies that we can compute $\lambda_1$ and $\lambda_2$ in $O(n^d)$ time and $O(n)$ space.
By further combining this with what we have in the previous section, we can finally conclude the following.
\begin{theorem} \label{thm-comp}
	One can compute the exact value of $\textnormal{comp}_\mathcal{S}$ in $O(n^d)$ time.
\end{theorem}

\bibliography{my_bib}

\begin{thebibliography}{10}

\bibitem{agarwal2013nearest}
Pankaj~K Agarwal, Boris Aronov, Sariel Har-Peled, Jeff~M Phillips, Ke~Yi, and
  Wuzhou Zhang.
\newblock Nearest neighbor searching under uncertainty ii.
\newblock In {\em Proceedings of the 32nd symposium on Principles of database
  systems}, pages 115--126. ACM, 2013.

\bibitem{agarwal2012range}
Pankaj~K Agarwal, Siu-Wing Cheng, and Ke~Yi.
\newblock Range searching on uncertain data.
\newblock {\em ACM Transactions on Algorithms (TALG)}, 8(4):43, 2012.

\bibitem{agarwal2014convex}
Pankaj~K Agarwal, Sariel Har-Peled, Subhash Suri, Hakan Y{\i}ld{\i}z, and
  Wuzhou Zhang.
\newblock Convex hulls under uncertainty.
\newblock In {\em Algorithms-ESA 2014}, pages 37--48. Springer, 2014.

\bibitem{agarwal2016range}
Pankaj~K Agarwal, Nirman Kumar, Stavros Sintos, and Subhash Suri.
\newblock Range-max queries on uncertain data.
\newblock In {\em Proceedings of the 35th ACM SIGMOD-SIGACT-SIGAI Symposium on
  Principles of Database Systems}, pages 465--476. ACM, 2016.

\bibitem{Edelsbrunner:1986:topological_sweep}
H.~Edelsbrunner and L.J. Guibas.
\newblock Topologically sweeping an arrangement.
\newblock In {\em Proceedings of the Eighteenth Annual ACM Symposium on Theory
  of Computing}, STOC '86, pages 389--403. ACM, 1986.

\bibitem{fink2016hyperplane}
Martin Fink, John Hershberger, Nirman Kumar, and Subhash Suri.
\newblock Hyperplane separability and convexity of probabilistic point sets.
\newblock In {\em Proceedings of the thirty-second annual symposium on
  Computational geometry}. ACM, 2016.

\bibitem{huang2015approximating}
Lingxiao Huang and Jian Li.
\newblock Approximating the expected values for combinatorial optimization
  problems over stochastic points.
\newblock In {\em International Colloquium on Automata, Languages, and
  Programming}, pages 910--921. Springer, 2015.

\bibitem{huang2014epsilon}
Lingxiao Huang, Jian Li, Jeff~M Phillips, and Haitao Wang.
\newblock $epsilon$-kernel coresets for stochastic points.
\newblock {\em arXiv preprint arXiv:1411.0194}, 2014.

\bibitem{jorgensen2011geometric}
Allan J{\o}rgensen, Maarten L{\"o}ffler, and Jeff~M Phillips.
\newblock Geometric computations on indecisive points.
\newblock In {\em Workshop on Algorithms and Data Structures}, pages 536--547.
  Springer, 2011.

\bibitem{kamousi2011stochastic2}
Pegah Kamousi, Timothy~M Chan, and Subhash Suri.
\newblock Stochastic minimum spanning trees in euclidean spaces.
\newblock In {\em Proceedings of the twenty-seventh annual symposium on
  Computational geometry}, pages 65--74. ACM, 2011.

\bibitem{kamousi2014closest}
Pegah Kamousi, Timothy~M Chan, and Subhash Suri.
\newblock Closest pair and the post office problem for stochastic points.
\newblock {\em Computational Geometry}, 47(2):214--223, 2014.

\bibitem{li2015expected}
Chao Li, Chenglin Fan, Jun Luo, Farong Zhong, and Binhai Zhu.
\newblock Expected computations on color spanning sets.
\newblock {\em Journal of Combinatorial Optimization}, 29(3):589--604, 2015.

\bibitem{loffler2009data}
Maarten L{\"o}ffler.
\newblock Data imprecision in computational geometry.
\newblock 2009.

\bibitem{loffler2010largest}
Maarten L{\"o}ffler and Marc van Kreveld.
\newblock Largest and smallest convex hulls for imprecise points.
\newblock {\em Algorithmica}, 56(2):235--269, 2010.

\bibitem{suri2014most}
Subhash Suri and Kevin Verbeek.
\newblock On the most likely voronoi diagramand nearest neighbor searching.
\newblock In {\em International Symposium on Algorithms and Computation}, pages
  338--350. Springer, 2014.

\bibitem{suri2013most}
Subhash Suri, Kevin Verbeek, and Hakan Y{\i}ld{\i}z.
\newblock On the most likely convex hull of uncertain points.
\newblock In {\em Algorithms--ESA 2013}, pages 791--802. Springer, 2013.

\bibitem{xue2016colored}
Jie Xue and Yuan Li.
\newblock Colored stochastic dominance problems.
\newblock {\em arXiv preprint arXiv:1612.06954}, 2016.

\bibitem{xue2016stochastic}
Jie Xue and Yuan Li.
\newblock Stochastic closest-pair problem and most-likely nearest-neighbor
  search in tree spaces.
\newblock {\em arXiv preprint arXiv:1612.04890}, 2016.

\bibitem{xue2016separability}
Jie Xue, Yuan Li, and Ravi Janardan.
\newblock On the separability of stochastic geometric objects, with
  applications.
\newblock In {\em Proceedings of the thirty-second annual symposium on
  Computational geometry}. ACM, 2016.

\end{thebibliography}
\newpage

\appendix
{\huge \bf \noindent Appendix}
\section{A simple 2-approximation of $\text{diam}_\mathcal{S}$} \label{app-2apx}
We describe a very simple $(n,d)$-polynomial-time 2-approximation of $\text{diam}_\mathcal{S}$, i.e., the expected diameter of a SCH of $\mathcal{S}$ (see Sec.~\ref{sec-prelim} and the beginning of Sec.~\ref{sec-diam} for the formal definition of $\text{diam}_\mathcal{S}$).
Given a set of points in $\mathbb{R}^d$, the diameter of their convex hull is just the distance between the farthest-pair of points (\textit{farthest-pair distance} hereafter).
Therefore, it suffices to approximate the expected farthest-pair distance of $\mathcal{S}$.
Recall the following well-known fact.
\begin{fact}
	For a set $X$ of points in a metric space and any $x \in X$, the farthest-pair distance of $X$ is at most $2\textnormal{dist}(x,y)$, where $y \in X$ is the point farthest from $x$.
\end{fact}
\textit{Proof.}
Assume the farthest-pair distance of $X$ is $\text{dist}(p,q)$ for some $p,q \in X$.
Then $2\text{dist}(x,y) \geq \text{dist}(x,p)+\text{dist}(x,q) \geq \text{dist}(p,q)$.
\hfill $\Box$ 
\medskip

By using the above fact, it is straightforward to obtain a 2-approximation algorithm for computing the expected farthest-pair distance of $\mathcal{S}$.
Suppose $\mathcal{S}=(S,\pi)$ with $S = \{a_1,\dots,a_n\}$.
If $R \subseteq S$ is a realization of $\mathcal{S}$, we consider the point in $R$ with the smallest index, say $a_i$, and the point in $R$ that is farthest from $a_i$, say $a_j$ (assume the points in $S$ have distinct distances from $a_i$).
Then $\text{dist}(a_i,a_j)$ is a 2-approximation of the farthest-pair distance of $R$.
We call $(a_i,a_j)$ the \textit{critical pair} of $R$.
Define $E_{i,j}$ as the event that a realization $R$ of $\mathcal{S}$ has critical pair $(a_i,a_j)$.
An 2-approximation of $\text{diam}_\mathcal{S}$ can be simply computed as

\begin{equation*}
\text{diam}^*_\mathcal{S} = \sum_{i=1}^{n} \sum_{j=1}^{n} \Pr[E_{i,j}] \cdot \text{dist}(a_i,a_j).
\end{equation*}
\normalsize
Note that $\Pr[E_{i,j}] = \pi(a_i) \cdot \pi(a_j) \cdot \prod_{k \in I_{i,j}} (1-\pi(a_k))$ where $I_{i,j} = \{k: k<i \text{ or } \text{dist}(a_i,a_j)<\text{dist}(a_i,a_k)\}$.
Therefore, a 2-approximation of $\text{diam}_\mathcal{S}$ can be computed in $(n,d)$-polynomial-time.
(More generally, one can compute a 2-approximation of the expected farthest-pair distance in any metric space in polynomial time, as long as the distance function can be computed in polynomial time.)

\section{\#P-hardness of the expected-diameter problem} \label{app-hard}

We prove the \#P-hardness of computing $\text{diam}_\mathcal{S}$ when the dimension $d$ is not assumed to be fixed (see Sec.~\ref{sec-diam} for the definition of $\text{diam}_\mathcal{S}$).
This extends a result in \cite{huang2015approximating} which states that computing the expected farthest-pair distance of a stochastic dataset in a (general) metric space is \#P-hard.

\begin{figure}[htpb]
	\centering
	\includegraphics[]{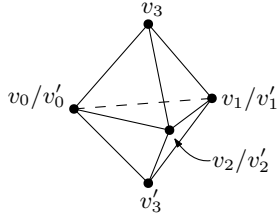}
	\caption{A regular double-simplex}
	\label{figure:regular double simplex}
\end{figure}

Our reduction is from the problem of counting independent sets of a graph, which is a well-known \#P-hard problem.
\begin{lemma} \label{lem-double}
	For an integer $k > 0$, there exists two positive real numbers $\alpha_k,\beta_k$ with $\alpha_k < \beta_k$ and a map $f:\{0,1,\dots,k,k+1\} \rightarrow \mathbb{R}^k$ such that $\textnormal{dist}(f(i),f(j)) = \alpha_k$ for any $i \neq j$ except $\textnormal{dist}(f(k),f(k+1)) = \textnormal{dist}(f(k+1),f(k)) = \beta_k$.
\end{lemma}
\textit{Proof.}
Let $\Delta$ be a regular $k$-simplex (i.e., a $k$-simplex with edges of length 1) with vertices $v_0,\dots,v_k$, $\Delta'$ be another regular $k$-simplex with vertices $v_0',\dots,v_k'$.
We form a \textit{regular double-simplex} by identically gluing the face $(v_0,\dots,v_{k-1})$ of $\Delta$ with the face $(v_0',\dots,v_{k-1}')$ of $\Delta'$ (see Figure~\ref{figure:regular double simplex}).
Clearly, this double-simplex can be (isometrically) embedded into $\mathbb{R}^k$ via an embedding map $\sigma$.
Now we define $f(k) = \sigma(v_k)$, $f(k+1) = \sigma(v_k')$, and $f(i) = \sigma(v_i) = \sigma(v_i')$ for all
$i \in \{0,\dots,k-1\}$.
By taking $\alpha_k = \text{dist}(f(0),f(1))$ and $\beta_k = \text{dist}(f(k),f(k+1))$, we complete the proof (the desired properties of $\alpha_k,\beta_k,f$ can be easily verified).
\hfill $\Box$
\medskip

\begin{lemma} \label{lem-embed}
	For a graph $G = (V,E)$, one can compute in polynomial time map $f:V \rightarrow \mathbb{R}^{|V|-1}$ such that
	
	\begin{equation*}
	\textnormal{dist}(f(u),f(v)) = \left\{
	\begin{array}{ll}
	\alpha & \textnormal{if } (u,v) \notin E, \\
	\beta & \textnormal{if } (u,v) \in E,
	\end{array}
	\right.
	\end{equation*}
	\normalsize
	for some $\alpha,\beta$ with $\alpha < \beta$.
\end{lemma}
\textit{Proof.}
Suppose $V = \{v_1,\dots,v_n\}$ and $E = \{e_1,\dots,e_m\}$.
Using Lemma~\ref{lem-double}, we find the real numbers $\alpha_{n-2},\beta_{n-2}$.
For each $e \in E$, let $g_e:V \rightarrow \mathbb{R}^{n-2}$ be a map such that

\begin{equation*}
\text{dist}(g_e(u),g_e(v)) = \left\{
\begin{array}{ll}
\alpha_{n-2} & \text{if } e \neq (u,v), \\
\beta_{n-2} & \text{if } e = (u,v).
\end{array}
\right.
\end{equation*}
\normalsize
Note that $g_e$ exists by Lemma~\ref{lem-double}.
We then define $g: V \rightarrow (\mathbb{R}^{n-2})^m$ by setting $g(v) = (g_{e_1}(v),\dots,g_{e_m}(v))$.
Let $\alpha = \sqrt{m} \cdot \alpha_{n-2}$ and $\beta = \sqrt{(m-1)\alpha_{n-2} + \beta_{n-2}}$.
It is easy to check that $\alpha<\beta$ and

\begin{equation*}
\text{dist}(g(u),g(v)) = \left\{
\begin{array}{ll}
\alpha & \text{if } (u,v) \notin E, \\
\beta & \text{if } (u,v) \in E.
\end{array}
\right.
\end{equation*}
\normalsize
To further construct $f$, we note that the image of $g$ consists only $n$ points, which should span a $(n-1)$-dim hyperplane in $(\mathbb{R}^{n-2})^m$.
If we (isometrically) identify this hyperplane with $\mathbb{R}^{n-1}$ and use $h:(\mathbb{R}^{n-2})^m \rightarrow \mathbb{R}^{n-1}$ to denote the projection map, $f:V \rightarrow \mathbb{R}^{n-1}$ is constructed as the composition $h \circ g$.
\hfill $\Box$
\medskip

With the above result in hand, we can now describe the reduction.
Given a graph $G = (V,E)$ with $V = \{v_1,\dots,v_n\}$, we first use Lemma~\ref{lem-embed} to compute the function $f:V \rightarrow \mathbb{R}^{n-1}$ and obtain $\alpha,\beta$.
Let $S$ be the $n$ points in the image of $f$.
We construct a stochastic dataset $\mathcal{S} = (S,\pi)$ by defining $\pi:S \rightarrow (0,1]$ as $\pi(a) = 0.5$ for all $a \in S$.
Now the subsets of $V$ are one-to-one corresponding to the realizations of $\mathcal{S}$.
By the construction of $f$, it is clear that a realization $R \subseteq S$ has a diameter $\text{diam}(R) = \alpha$ if $R$ corresponds to an independent set of $G$, and has a diameter $\text{diam}(R) = \beta$ otherwise.
Furthermore, every subset of $S$ occurs as a realization with an equal probability $2^{-n}$.
Hence, we immediately obtain the equation

\begin{equation*}
\text{diam}_\mathcal{S} = \beta + 2^{-n} \mathit{Ind}(G) \cdot (\alpha-\beta),
\end{equation*}
\normalsize
where $\mathit{Ind}(G)$ is the number of the independent sets of $G$.
In this way, counting the independent sets of $G$ is reduced to computing $\text{diam}_\mathcal{S}$, which implies the following hardness result.
\begin{theorem}
	Computing $\textnormal{diam}_\mathcal{S}$ is \#P-hard without assuming $d$ is fixed.
\end{theorem}

\section{Improving the algorithm for approximating $\text{diam}_\mathcal{S}$} \label{app-impdiam}
In this section, we show how to improve the runtime of our algorithm in Sec.~\ref{sec-algdiam} (see Sec.~\ref{sec-algdiam} for the notations used).
Fixing $p_1,p_2,p_3,p_4 \in S$, we show how to compute $\Pr[\psi]$ for all $\psi \in \varPsi_S$ of the form $\psi=(p_1,\dots,p_4,\cdot)$ in $O(n \log n)$ time.
As argued before, we may assume $p_1 \neq p_2$.
Let $r$ be the ray with initial point $p_2$ which goes through $p_1$, and $x$ be the point on $r$ which has distance $\textnormal{dist}(p_2,p_3)/2$ from $p_2$.
First, we determine a subset $A \subseteq S$ consisting of $p_4$ and all the points $a \in S$ satisfying $p_1 \prec a$ or $p_2 \prec_{p_1} a$ or $p_3 \prec_{p_2} a$ or $p_4 \prec_x a$.
It is clear that $\Pr[\psi]>0$ for $\psi=(p_1,\dots,p_4,p)$ only if $p \in S \backslash A$.
For each $p \in S \backslash A$, we denote by $B_p$ the set of all points $b \in S \backslash A$ with $p \prec_{p_4} b$.
By Lemma~\ref{lem-seqcond}, we have

\begin{equation} \label{eq-prpsi}
\Pr[\psi_p] = \left(\prod_{i=1}^{4}\pi(p_i) \cdot \prod_{a \in A} (1-\pi(a))\right) \cdot \left(\pi(p) \cdot \prod_{b \in B_p} (1-\pi(b))\right),
\end{equation}
\normalsize
where $\psi_p = (p_1,p_2,p_3,p_4,p)$.
Note that the left part of the above formula is independent of $p$ and thus only needs to be computed once.
To compute the right part efficiently, suppose $S \backslash A = \{c_1,\dots,c_r\}$.
We relabel these points such that $c_1 \prec_{p_4} \cdots \prec_{p_4} c_r$.
This can be done by sorting in $O(n \log n)$ time, or more precisely, $O(r \log r)$ time.
We then compute $\prod_{j=i}^{r} (1-\pi(c_j))$ for all $i \in \{1,\dots,r\}$ (note that this can be done in linear time).
With this in hand, we consider each $p \in S \backslash A$.
We must have $p = c_i$ for some $i \in \{1,\dots,r\}$.
In this case, the right part of Equation~\ref{eq-prpsi} is just $\pi(c_i) \cdot \prod_{j=i+1}^{r} (1-\pi(c_j))$ and hence can be computed in constant time.
Therefore, we can compute $\Pr[\psi_p]$ for all $p \in S \backslash A$ in linear time.
Including the time for sorting, this gives us an $O(n^5 \log n)$-time 1.633-approximation algorithm for computing $\text{diam}_\mathcal{S}$.

\section{Improving the algorithm for approximating $\text{wid}_\mathcal{S}$} \label{app-impwid}
In this section, we show how to improve the runtime of our algorithm in Sec.~\ref{sec-algwid} (see Sec.~\ref{sec-algwid} for the notations used).
We enumerate all $\Delta \in \varGamma_S^d$ by considering their vertex lists.
Fixing $d$ (distinct) points $v_0,\dots,v_{d-1} \in S$, we show how to compute $\Pr[\Delta]$ for all $\Delta \in \varGamma_S^d$ whose vertex lists are of the form $(v_0,\dots,v_{d-1},\cdot)$ in $O(n \log n)$ time.
First, we determine a subset $V \subseteq S \backslash \{v_0,\dots,v_{d-1}\}$ consisting of all $v \in S \backslash \{v_0,\dots,v_{d-1}\}$ such that $(v_0,\dots,v_{d-1},v)$ is the vertex list of the $d$-simplex whose vertices are $v_0,\dots,v_{d-1},v$.
Clearly, this step can be completed in linear time by enumerating all $v \in S \backslash \{v_0,\dots,v_{d-1}\}$ and verifying for each $v$ whether $v \in V$.
If $V = \emptyset$, we are done because there is no $\Delta \in \varGamma_S^d$ whose vertex list is of the form $(v_0,\dots,v_{d-1},\cdot)$.
So suppose $V \neq \emptyset$.
For $i \in \{0,\dots,d-1\}$, we denote by $E_i$ be the $i$-dim hyperplane in $\mathbb{R}^d$ through $v_0,\dots,v_{i}$.
We then compute a subset $A \subset S$ consisting of all $a \in S$ such that $v_0 \prec a$ or $v_{i+1} \prec_{E_i} a$ for some $i \in \{0,\dots,d-2\}$.
Now for any $v \in V$, we denote by $B_v$ the set of all $b \in S \backslash A$ such that $v \prec_{E_{d-1}} b$.
By Lemma~\ref{lem-conditions}, we have

\begin{equation} \label{eq-prdelta}
\Pr[\Delta_v] = \left( \prod_{i=0}^{d-1} \pi(v_i) \cdot \prod_{a \in A} (1-\pi(a)) \right) \cdot \left( \pi(v) \cdot \prod_{b \in B_v} (1-\pi(b)) \right),
\end{equation}
\normalsize
where $\Delta_v$ is the $d$-simplex with vertices $v_0,\dots,v_{d-1},v$.
Note that the left part of the above formula is independent of $v$ and thus only needs to be computed once.
To compute the right part efficiently, suppose $S \backslash A = \{c_1,\dots,c_r\}$.
We relabel these points such that $c_1 \prec_{E_{d-1}} \cdots \prec_{E_{d-1}} c_r$.
This can be done by sorting in $O(n \log n)$ time, or more precisely, $O(r \log r)$ time.
We then compute $\prod_{j=i}^{r} (1-\pi(c_j))$ for all $i \in \{1,\dots,r\}$ (note that this can be done in linear time).
With this in hand, we consider each $v \in V$.
Since $V \subseteq S \backslash A$, we must have $v = c_i$ for some $i \in \{1,\dots,r\}$.
In this case, the right part of Equation~\ref{eq-prdelta} is just $\pi(c_i) \cdot \prod_{j=i+1}^{r} (1-\pi(c_j))$ and hence can be computed in constant time.
Therefore, we can compute $\Pr[\Delta_v]$ for all $v \in V$ in linear time.
Including the time for sorting, this gives us an $O(n^{d+1} \log n)$ time algorithm for computing $\text{wid}_\mathcal{S}^*$, i.e., approximating $\text{wid}_\mathcal{S}$ within a constant factor.


\section{Computing the $\mathcal{S}$-statistics for $\mathcal{E}$} \label{app-reptstat}
In this section, we describe an algorithm for computing the $\mathcal{S}$-statistics for $\mathcal{E}$ in $O(n^d)$ time and $O(n)$ space (see Sec.~\ref{sec-special} for the definition of $\mathcal{S}$-statistics and $\mathcal{E}$).
Suppose $S = \{a_1,\dots,a_n\}$.
Then every hyperplane $E \in \mathcal{E}$ can be uniquely represented as a $d$-tuple $(a_{i_1},\dots,a_{i_d})$ where $a_{i_1},\dots,a_{i_d}$ are the points on $E$ and $i_1<\dots<i_d$.
We first describe an algorithm using $O(n^d \log n)$ time and $O(n)$ space.
Fixing $d-1$ points $a_{i_1},\dots,a_{i_{d-1}} \in S$ with $i_1<\cdots<i_{d-1}$, we show how to report, in $O(n \log n)$ time and $O(n)$ space, the $\mathcal{S}$-statistics of all hyperplanes (i.e., lines) in $\mathcal{E}$ which are represented as the form $(a_{i_1},\dots,a_{i_{d-1}},\cdot)$.
Define $Y$ as the $(d-2)$-dim hyperplane in $\mathbb{R}^d$ spanned by $a_{i_1},\dots,a_{i_{d-1}}$.
Let $Z$ be the (unique) \textit{vertical} $(d-1)$-dim hyperplane containing $Y$ (by ``vertical'' we mean that $Z$ is perpendicular to the hyperplane $x_d = 0$), and $\mathcal{E}' \subseteq \mathcal{E}$ be the sub-collection consisting of all hyperplanes in $\mathcal{E}$ which contain $Y$.
Note that $|\mathcal{E}'| = n-d+1$.
We then sort the hyperplanes in $\mathcal{E}'$ in the \textit{rotation order} around $Y$, that is, we assign to each hyperplane $E \in \mathcal{E}'$ a key value equal to the rotation angle from $Z$ to $E$ (the rotation is taken around $Y$ with a fixed direction), and sort the lines by their key values.
Assume $E_1,\dots,E_{n-d+1}$ is the sorted list.
Observe that $\text{stat}(E_{j+1})$ can be computed in constant time if $\text{stat}(E_j)$ is already in hand, basically because the points on each side of $E_{j+1}$ are almost the same as those on each side of $E_j$ except two points.
By this observation, we may compute the $\mathcal{S}$-statistics of $E_1,\dots,E_{n-d+1}$ in $O(n)$ time.
Once $\text{stat}(E_j)$ is computed, we report it if $E_j$ is represented as the form $(a_{i_1},\dots,a_{i_{d-1}},\cdot)$.
In this way, we obtain an $O(n^d \log n)$-time and $O(n)$-space algorithm.
To shave off the $\log n$ factor in the time bound, we need to further apply the techniques of duality and topological sweep \cite{Edelsbrunner:1986:topological_sweep}.
This approach heavily relies on an idea in \cite{fink2016hyperplane} (which was used to improve the algorithm for computing the separability-probability), so here we only provide a sketch.
Instead of fixing $d-1$ points, we fix $d-2$ points $a_{i_1},\dots,a_{i_{d-2}} \in S$ with $i_1<\cdots<i_{d-2}$, and want to report, in $O(n^2)$ time and $O(n)$ space, $\text{stat}(E)$ for all $E \in \mathcal{E}$ which are represented as the form $(a_{i_1},\dots,a_{i_{d-2}},\cdot,\cdot)$.
Note that if this can be done, we immediately obtain an $O(n^d)$-time and $O(n)$-space algorithm.
Consider the point-set $S$ in the dual space of $\mathbb{R}^d$.
Every point $a_i \in S$ is dual to a $(d-1)$-dim hyperplane $a_i^*$ in the dual space.
Furthermore, a $(k-1)$-dim hyperplane spanned by $k$ (distinct) points $a_{j_1},\dots,a_{j_k} \in S$ is dual to a $(d-k)$-dim hyperplane in the dual space, which is in fact the intersection of $a_{j_1}^*,\dots,a_{j_k}^*$.
Let $D$ be the $(d-3)$-dim hyperplane spanned by $a_{i_1},\dots,a_{i_{d-2}}$, which is dual to a 2-dim hyperplane (i.e., a plane) $D^*$ in the dual space.
For each $a_i \in S \backslash \{a_{i_1},\dots,a_{i_{d-2}}\}$, the intersection of $a_i^*$ and $D^*$ is a line in $D^*$ (which should be the dual of the $(d-2)$-dim hyperplane spanned by $a_{i_1},\dots,a_{i_{d-2}},a_i$).
These $n-d+2$ lines form a line arrangement in $D^*$.
Suppose $l_i^*$ is the line corresponding to $a_i$.
In the line arrangement, there are $n-d+1$ intersection points on $l_i^*$, each of which is the dual of a hyperplane through $a_{i_1},\dots,a_{i_{d-2}},a_i$ in the original space.
The order of these intersection points appearing on $l_i^*$ is just the rotation order of the corresponding hyperplanes in the original space.
Therefore, if these intersection points are already sorted, we can compute the $\mathcal{S}$-statistic of each of the corresponding hyperplanes in amortized $O(1)$ time.
But we cannot use sorting, as it takes $O(n \log n)$ time per line and we have $O(n)$ lines in the arrangement.
Instead, we use topological sweep to visit the intersection points in the arrangement.
In the process of topological sweep, the intersection points on each line is visited in order along the line (though not consecutively).
When the first intersection point on a line is visited, we use brute-force to compute the $\mathcal{S}$-statistic of the corresponding hyperplane in $O(n)$ time.
Then when we go to the next intersection point on the line, we can compute the $\mathcal{S}$-statistic of the corresponding hyperplane in constant time from the $\mathcal{S}$-statistic of the hyperplane corresponding to the previous intersection point.
Once a $\mathcal{S}$-statistic is computed, we report it if the hyperplane is represented as the form $(a_{i_1},\dots,a_{i_{d-2}},\cdot,\cdot)$.
The topological sweep takes $O(n^2)$ time and $O(n)$ space.
Thus, we obtain an algorithm computing the $\mathcal{S}$-statistics for $\mathcal{E}$ in $O(n^d)$ time and $O(n)$ space.

\end{document}